\documentclass[fleqn,usenatbib]{mnras}

\usepackage[T1]{fontenc}
\usepackage{ae,aecompl}
\pdfoutput=1

\usepackage[pdftex]{graphicx}	
\usepackage{amsmath}	
\usepackage{amssymb}	
\usepackage{xcolor}

\usepackage{newtxtext,newtxmath}

\title[\textit{Swift} J1858.6$-$0814]{Long term radio monitoring of the neutron star X-ray binary \textit{Swift} J1858.6$-$0814}

\author[L. Rhodes et al.]{
L. Rhodes,$^{1,2}$\thanks{E-mail: lauren.rhodes@physics.ox.ac.uk}
R. P. Fender,$^{1,3}$
S. Motta,$^{4}$
J. van den Eijnden,$^{1}$
D. R. A. Williams,$^{5}$
\newauthor
J. Bright,$^{6}$
G. R. Sivakoff,$^{7}$
\\
$^{1}$ Astrophysics, University of Oxford, Denys Wilkinson Building, Keble Road, Oxford, OX1 3RH, UK\\
$^{2}$ Max Planck Institute f\"{u}r Radioastronomie, Auf dem H\"{u}gel, Bonn 53121, Germany\\
$^{3}$ Department of Astronomy, University of Cape Town, Private Bag X3, Rondebosch 7701, South Africa\\
$^{4}$ INAF, Osservatorio Astronomico di Brera, via Brera 28, 20121 Milano, Italy\\
$^{5}$ Jodrell Bank Centre for Astrophysics, School of Physics and Astronomy, The University of Manchester, Manchester, M13 9PL, UK\\
$^{6}$ Department of Astronomy, University of California, Berkeley, CA 94720-3411, USA\\
$^{7}$ Department of Physics, University of Alberta, CCIS4-181 Edmonton, AB TG6 2E1, Canada
}

\date{Accepted XXX. Received YYY; in original form ZZZ}

\pubyear{2015}

\begin{document}

\label{firstpage}
\pagerange{\pageref{firstpage}--\pageref{lastpage}}
\maketitle

\begin{abstract}
We present the results of our long term radio monitoring campaign at 1.3GHz (MeerKAT) and 15.5\,GHz (Arcminute Microkelvin Imager - Large Array, AMI-LA) for the outburst of the recently discovered neutron star X-ray binary \textit{Swift} J1858.6$-$0814. Throughout the outburst, we observe radio emission consistent with a quasi-persistent, self-absorbed jet. In addition, we see two flares at MJD 58427 and 58530. The second flare allows us to place constraints on the magnetic field and minimum energy of the jet at 0.2\,G and 5$\times$10\textsuperscript{37}\,erg, respectively. We use the multi-frequency radio data in conjunction with data from \textit{Swift}-BAT to place \textit{Swift} J1858.6-0814 on the radio/X-ray correlation. We find that the quasi-simultaneous radio and BAT data makes \textit{Swift} J1858.6-0814 appears to bridge the gap in the radio/X-ray plane between atoll and Z sources. Furthermore, AMI-LA observations made whilst \textit{Swift} J1858.6-0814 was in the soft state have allowed us to show that the radio emission during the soft state is quenched by at least a factor of four.  
\end{abstract}




\section{Introduction}
Low mass X-ray binaries (XRBs) are systems which contain a low mass star and a compact object - either a neutron star or stellar mass black hole - orbiting each other. As the stellar companion evolves and fills its Roche Lobe, matter is transferred to the compact object via an accretion disc. Most black hole and some neutron star XRBs are transient: instabilities in the disc cause an increase in accretion rate, causing the system to go into `outburst' \citep{Narayan1995, 2001NewAR..45..449L}. 

Neutron star XRBs are classified based on their timing properties and the paths they track on a colour-colour diagram \citep[a plot to study the apparent X-ray colour of XRB by comparing the number of counts in harder and softer X-ray bands e.g.][]{1989ESASP.296..203V}. All neutron star XRBs are thought to have low magnetic fields \citep[$\sim$10\textsuperscript{9}\,G, e.g. ][]{2016MNRAS.461.4049D, 2019ApJ...873...99L}. They appear to fall into two separate classes: Atoll and Z sources. Studies of XTE J1701-462 have shown that the class apparent separation is a result in changes in mass accretion rate \citep{2010ApJ...719..201H}.

Atoll sources share some characteristics with black hole XRBs: they have similar X-ray spectra and timing properties \citep{1994ApJS...92..511V, 2006ARA&A..44...49R, 2014MNRAS.443.3270M}. They differ in their radio properties where atoll sources are 27 times less radio luminous \citep{2018MNRAS.478L.132G}. Z sources evolve much more quickly. They traverse their colour-colour diagram in hours-days, thought to be caused by variations in mass transfer rate, which is usually near the Eddington limit \citep{1989A&A...225...79H}. Z sources are more X-ray and radio luminous than atoll sources. Both the X-ray and radio emission varies rapidly along with transient ejections \cite[e.g.][]{ 2001ApJ...558..283F, 2004Natur.427..222F}.

At the beginning of the outburst, black hole and atoll neutron star systems are in a hard X-ray photon dominated, low luminosity state, where the X-ray spectrum is parameterised by a power law index ($\Gamma$). The emission is thought to be produced by inverse Compton scattering of disc photons to higher energies. This is referred to as the hard state. As the outburst evolves, systems enter a higher luminosity (higher accretion rate) state. Here, the X-ray spectrum is dominated by soft photons produced by the disc of some black body temperature, which we call the soft state. Sometimes the hard and soft states are referred to the extreme island and banana states, respectively, for atoll sources. Before returning to quiescence, these systems return to the hard state in a hysteresis pattern \citep[][for a recent review]{2009MNRAS.396.1370F, 2016ASSL..440...61B}. In all states, neutron star XRB spectra may also contain an additional component originating from the neutron star's surface \citep{2021arXiv210800729M}.

XRBs in the hard state have been observed to have associated flat or inverted radio emission \citep[flux density, S$_{\nu} \propto \nu^{\alpha}$ where $\alpha \geq 0$,][]{Fender2001, 2003MNRAS.342L..67M}. The radio emission is thought to be a result of the accretion process launching compact steady jets \citep{Fender2004}. Relativistic electrons gyrate along the magnetic field lines in the jet producing synchrotron emission which is the dominant emission mechanism at radio frequencies.


Upon transitioning from the hard to soft state, radio emission in black hole XRBs is quenched by at least 3.5 orders of magnitude \citep{1999ApJ...519L.165F, 2011ApJ...739L..19R, 2019ApJ...883..198R, 2021MNRAS.504..444C}. During the transition between states, radio flares are often observed along with transient ejections: blobs of plasma that are ejected from the system and propagate into the surrounding ISM \citep{ 1994Natur.371...46M, 1995Natur.375..464H, 2016ASSL..440...61B}. It appears that neutron star XRBs have a more complex jet quenching process. Some systems have shown clear evidence of strong quenching e.g. Aquila X-1  and 1RXS J180408.9-342058 \citep[][ respectively]{2010ApJ...716L.109M, 2017MNRAS.470.1871G}. Instead, 4U 1820 - 30 shows some reduction in radio flux density and change in spectral index consistent with a quenching of the compact jet and possible launch of transient ejecta \citep{2004MNRAS.351..186M, 2021MNRAS.508L...6R}.

\subsection{Swift J1858.6-0814}
\label{sub:J1858}

\textit{Swift} J1858.6-0814 (hereafter J1858) is a newly discovered XRB that was first reported as going into outburst by the Neil Gehrels \textit{Swift} Observatory (hereafter \textit{Swift}) - Burst Alert Telescope (BAT) on MJD 58416 (2018 October 25) \citep{Krimm2018}. The reported position was RA: 18h 58m 34.96s, dec: -08d 14' 16.4" with an uncertainty of 2.2\,arcseconds \citep{2018ATel12160....1K}.

A number of multi-wavelength monitoring campaigns of this outburst have been conducted, including radio (presented in this work), optical \citep{2020ATel13719....1S} and X-ray \citep{2020MNRAS.496.4127V, 2021MNRAS.503.5600B}. All sky X-ray monitors also observed this source \citep{2009PASJ...61..999M, Krimm2013}\footnote{https://swift.gsfc.nasa.gov/results/transients/weak/SWIFTJ1858.6-0814/, http://maxi.riken.jp/top/lc.html}. There have also been many targeted, short term studies which provide in-depth, short timescale information on J1858. For the first $\sim$18\,months of the outburst, there were reports of J1858's X-ray spectra being best fit with a power law, indicative of the source being in the hard/extreme island state. However the power law index of the individual spectra varied massively between 0.22 and 2.2 \citep{Kennea2018, Bozzo2018}. There was also evidence of strong variability and soft flares on timescales of 10-100\,seconds along with variable absorption centred around N\textsubscript{H} = $2\times10^{21}$cm\textsuperscript{-2} \citep{2018ATel12158....1L, 2020ApJ...890...57H}. Despite the fact that a power law was the best spectral fit on many occasions, the varied behaviour has made it hard to conclusively say that J1858 was in the hard state, unlike many other black hole/ atoll XRBs \citep{2020ApJ...890...57H}.

\citet{2020ATel13536....1B} reported that J1858 had made a state transition to the soft state from at least MJD 58897 (after $\sim$18\,months in a hard state) evidenced by a significantly brighter soft X-ray flux with NICER (MJD 58897). Shorter, more details studies of J1858 with NICER and NuSTAR were performed after the state change. Firstly, they determined that the compact object is a neutron star via the detection of Type I X-ray bursts \citep{2020ATel13563....1B}. Type I X-ray bursts are thought to be evidence of photospheric radius expansion occurring when the X-ray flux observed corresponds to the Eddington luminosity. As a result these events can be considered to be standard candles \citep{1984ApJ...276L..41T, 1984ApJ...277L..57L}. The distance calculated from the detection of Type I X-ray bursts is 12.8$^{+0.7}_{-0.6}$kpc, \citet{2020MNRAS.499..793B} notes that the distance is  metallicity and inclination dependent where a higher hydrogen mass fraction and inclinations favours lower distances. Secondly, NICER detected the presence of X-ray eclipses allowing for the calculation of the orbital period ($\sim$0.83\,days) as well as implying that J1858 has a high inclination \citep[$>$75$^{\circ}$, ][]{2021MNRAS.503.5600B}. Such a high inclination could impact observing the inner regions of the accretion disk, which would explain the increased and highly variable absorption that has been observed \citep{2020ApJ...890...57H}.

Even with a state change, they report that the increased soft X-ray flux is not enough to conclude that J1858 has definitely transitioned into a canonical soft state. It may be that J1858 follows the established hard/soft state evolution as seen in other atoll systems \citep{2014MNRAS.443.3270M}, but the obscuration and variability observed at X-ray energies makes it difficult to conclusively determine J1858's long term X-ray state evolution.

Optical variability has also been observed \citep{2018ATel12197....1P} in addition to detections of optical P-Cygni profiles, a signature ascribed to out-flowing material \citep{2019ATel12881....1M}. Combining the optical and X-ray observations has led to comparisons to V404 Cygni and V4641 Sgr, which showed similar variability and absorption features \citep{2020MNRAS.499..793B, 2020ApJ...893L..19M}. \citet{2020MNRAS.496.4127V} also observed minute-timescale radio variability with the Karl G Jansky Very Large Array (VLA) and the Australia Telescope Compact Array (ATCA). In conjunction with multi-wavelength observations of J1858, we commenced a long term radio monitoring campaign with with the Arcminute Microkelvin Imager - Large Array (AMI-LA) and MeerKAT. Observations were triggered on MJD 58424 and MJD 58432 with AMI-LA and MeerKAT, respectively. The initial radio detection with AMI-LA was reported in \citet{2018ATel12184....1B}. 

In order to relate the radio and X-ray behaviour throughout the rest of the paper, we refer to the period between MJD 58416 and 58897 as the `hard state' and between MJD 58897 and 58928 as the `soft state' \citep{2020ATel13536....1B}. After MJD 58928, J1858 transitioned into quiescence \citep{2020ATel13719....1S}.

In Section \ref{section:obs}, we describe the observations and data reduction process for both the AMI-LA and MeerKAT as part of the ThunderKAT Large Survey Project \citep{Fender2017}. We combine this with \textit{Swift}-BAT data taken as part of the hard X-ray transient monitoring program \citep{Krimm2013}. In Section \ref{section:results}, we present the results of our observing campaign and in Section \ref{sec:discussion}, we place J1858 in the context of other neutron star XRBs. 

\section{Observations and Data Analysis}
\label{section:obs}

\subsection{Arcminute Microkelvin Imager - Large Array}
\label{subsec:AMI}

We started observing J1858 with the Arcminute Microkelvin Imager (AMI-LA) on MJD 58424 (2018 November 2) for 2 hours (8 days after the outburst began). The observations were made at 15.5\,GHz with a bandwidth of 5\,GHz, binned to 8 channels \citep{2008MNRAS.391.1545Z, 2018MNRAS.475.5677H}. All AMI-LA data were reduced using a custom pipeline: \textsc{reduce\_dc}, which flags the data, performs a primary calibration (using 3C 286), and applies phase corrections, using J1846$-$0651 as a secondary calibrator \citep{2013MNRAS.429.3330P}. J1846$-$0651 was observed for 60-90\,seconds every 10 minutes. The data were cleaned and imaged in \textsc{casa} \citep[Version 4.7.0][]{2007ASPC..376..127M}.

In the initial 2 hour observation, J1858 was detected with a flux density of 310$\pm$50$\mu$Jy/beam. 
In the first week, five observations were made. Following that, we observed J1858 once every one to two weeks for the rest of the outburst; 59 observations in total over 18 months. 

The colour map in Figure \ref{fig:maps} shows the central region of the AMI field. The source labelled \textit{1} is J1858. There is an additional, steady, $\sim$200$\mu$Jy/beam source within 70" of the phase centre, labelled \textit{2}. 

Some images of J1858 made with AMI-LA had an elliptical beam as a result of the source's equatorial declination. The elliptical beam produced issues in some epochs where the position angle of the AMI-LA clean beam is such that both source \textit{1} and \textit{2} were unresolvable. In three of the epochs where J1858 and the second source were unresolvable, we generated a point source model and performed a \textit{uv}-subtraction in \textsc{casa}, using the task \textit{uv-sub}, to subtract emission due to the secondary source (assuming a flux density of 200$\mu$Jy). The resulting flux density measurements of J1858 are denoted by unfilled black circles in Figure \ref{fig:lc} and \ref{fig:early}. Even after \textit{uv}-subtraction, there may be some contamination from the secondary source, if the secondary source is variable, hence the brighter than average flux densities in these epochs. We note that it was not possible to successfully perform \textit{uv}-subtraction for all epochs where the J1858 and source \textit{2} were unresolved. For these observations we report upper limits. AMI-LA observations ceased March 2020 due to the COVID-19 pandemic, our final epoch was on MJD 58924. Table \ref{tab:AMI_obs} gives a full list of all observing dates, durations, the flux density measurements and upper limits from the AMI-LA observing run.

\subsection{MeerKAT}

Monitoring of J1858 with MeerKAT was performed as part of the large survey project ThunderKAT \citep{Fender2017}. Weekly observations of J1858 with MeerKAT commenced on the MJD 58432 (2018 November 10), 16 days after the outburst was initially reported) and ceased on MJD 58488 (2019 January 5).  

Weekly monitoring restarted on MJD 58530 (2019 February 16), as J1858 was no longer sun constrained and new optical observations showed that the source was still in outburst, and continued until MJD 58632 (May 25) \citep{Rajwade2019}. We stopped regular monitoring after another series of three non-detections. Two further, isolated observations were made: the first (MJD 58700) was part of a multi-wavelength campaign \citep{CastroSegura2022}, the second was after a report of an X-ray state change \citep[MJD $\sim$58900,][]{2020ATel13455....1N, 2020ATel13563....1B}. 

The MeerKAT observations were made at a central frequency of 1.28\,GHz with a bandwidth of 856\,MHz split into 4096 channels. Each observation consisted of 10 minutes observing the bandpass and flux calibrator, PKS B1934$-$638 (J1939$-$6342), 2 minutes observing the phase calibrator, PKS J1911$-$2006, both before and after J1858, combined with 15 minutes on J1858. Reduction of the MeerKAT data was performed using a set of python scripts, \texttt{OxKAT}, allowing for semi-automatic processing \citep{2020ascl.soft09003H}. Persistent RFI was removed from the calibrator fields before performing bandpass calibration and flux density scaling using J1939$-$6342. The complex gains were solved for on both J1939$-$634 and J1911$-$2006 and applied to the target field, which was flagged using \textsc{tricolour}\footnote{https://github.com/ska-sa/tricolour}. 
Imaging was performed with \textsc{wsclean} \citep[Version 2.5,][]{2014MNRAS.444..606O}, using a Briggs weighting with a robust parameter of -0.7. We also performed a single round of phase-only self calibration after which the target field was re-imaged. 

\begin{figure}
    \centering
    \includegraphics[width=\columnwidth]{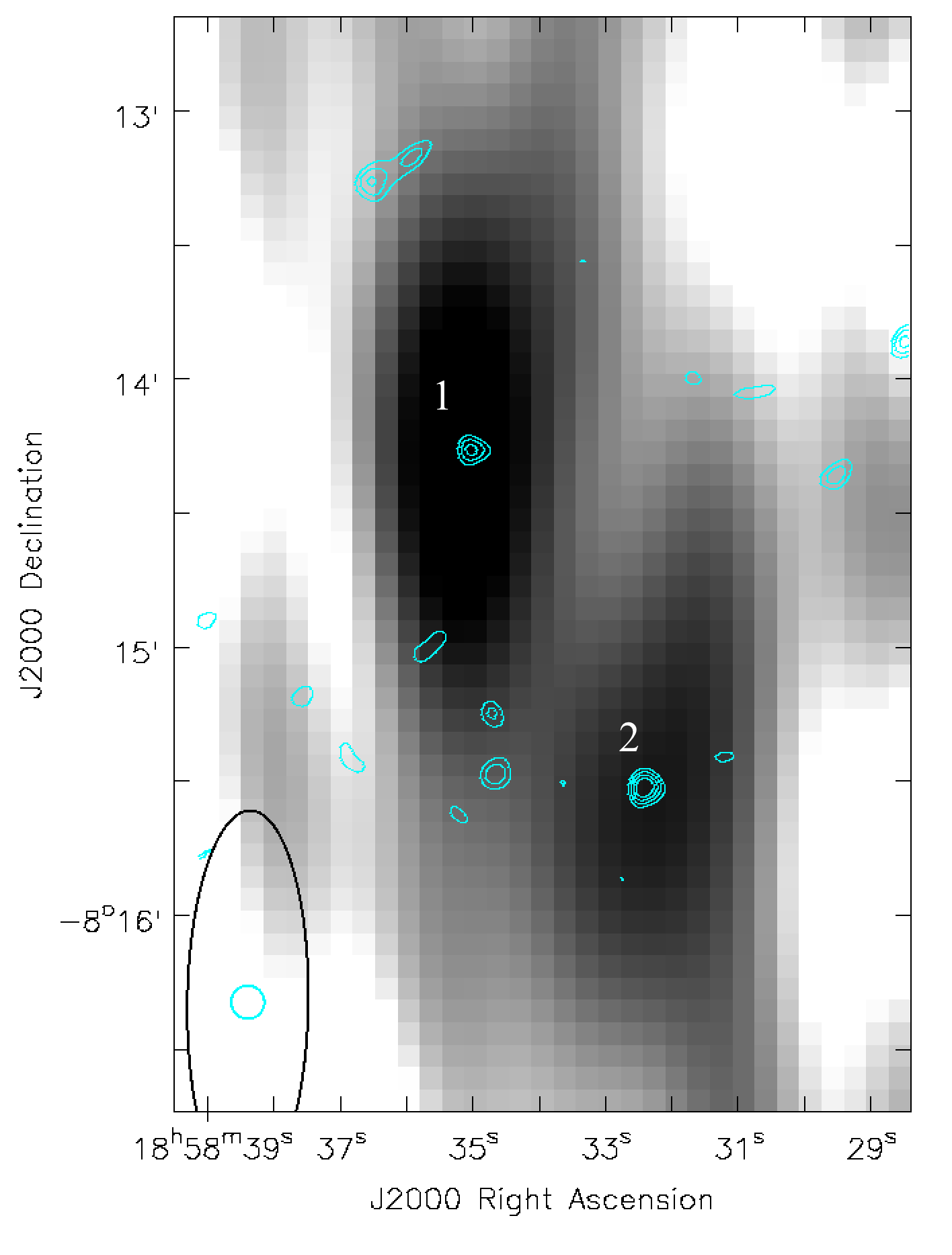}
    \caption{The J1858 field at radio frequencies. The background colour map shows an AMI observation in which J1858 and the south-west source are separated. At some epochs, the synthesised beam position angle resulted in J1858 (source 1) and source 2 being blended in the image. The cyan contours show an example of the MeerKAT field. The contours are at 3, 4, 5 and 6 sigma times an rms noise of 22$\mu$Jy. The source labelled (1) is J1858, also labelled is the second source (2) which has created issues in some AMI observations.} 
    \label{fig:maps}
\end{figure}

The contours in Figure \ref{fig:maps} show the central region of the MeerKAT J1858 field. Compared to the AMI field, the sources labelled \textit{1} and \textit{2} in Figure \ref{fig:maps} are clearly resolved in MeerKAT image. Table \ref{tab:MeerKAT_obs} shows a full list of the observation made with MeerKAT, the flux density and uncertainty for each epoch. 

\subsection{\textit{Swift}-BAT}
\textit{Swift}-BAT first triggered on the outburst of J1858 on MJD 58416 \citep{Krimm2018}. Our first data point is 6 days later, on MJD 58422. BAT tracked the X-ray emission from J1858 between 15 and 50\,keV, until MJD 58935 when the source returned to quiescence \citep{2020ATel13719....1S}. We obtained the data from the \textit{Swift}-BAT hard transient monitoring program. The third panel in Figures \ref{fig:lc} and \ref{fig:flare_zoom} shows the BAT light curves. The flux levels for J1858 are very faint, therefore, we only plot the epochs where we obtain at least a 3$\sigma$ detection which we calculate by assuming that the flux uncertainty is representative of the rms noise. A 3$\sigma$ detection would be at least as bright as three times the uncertainty on the detection. 

In order to compare J1858 to other hard state XRBs, we converted the 15-50\,keV count rate to a 1-10\,keV luminosity. First, we extrapolated the 15-50\,keV flux to a 1-10\,keV unabsorbed flux using a range of photon indices ($\Gamma$) within \texttt{WebPIMMS}\footnote{https://heasarc.gsfc.nasa.gov$/$cgi$-$bin$/$Tools$/$w3pimms$/$w3pimms.pl}. The use of 15-50\,keV count rates is discussed further in Section \ref{sec:discussion}. We used two $\Gamma$ values: 1.4 and 2.2 \citep{Kennea2018, Bozzo2018}. X-ray spectra of J1858 have been fit with a broader range of $\Gamma$, between 0.22 and 2.2 however anything below 1.4 becomes nonphysical due to the nature of the Comptonization of soft photons from the disc \citep{2007A&ARv..15....1D}. We note that the unabsorbed flux is used because at low energies the X-ray photons are susceptible to absorption or scattering. Therefore, by extrapolating from higher X-ray energies we obtain a better estimate of the accretion flow. From there, the unabsorbed fluxes were converted to luminosities which are used in Figure \ref{fig:lr-lx}. 

\subsection{MAXI/GSC}
\label{sub:MAXI}

\begin{figure*}
    \centering
    \includegraphics[width=\textwidth]{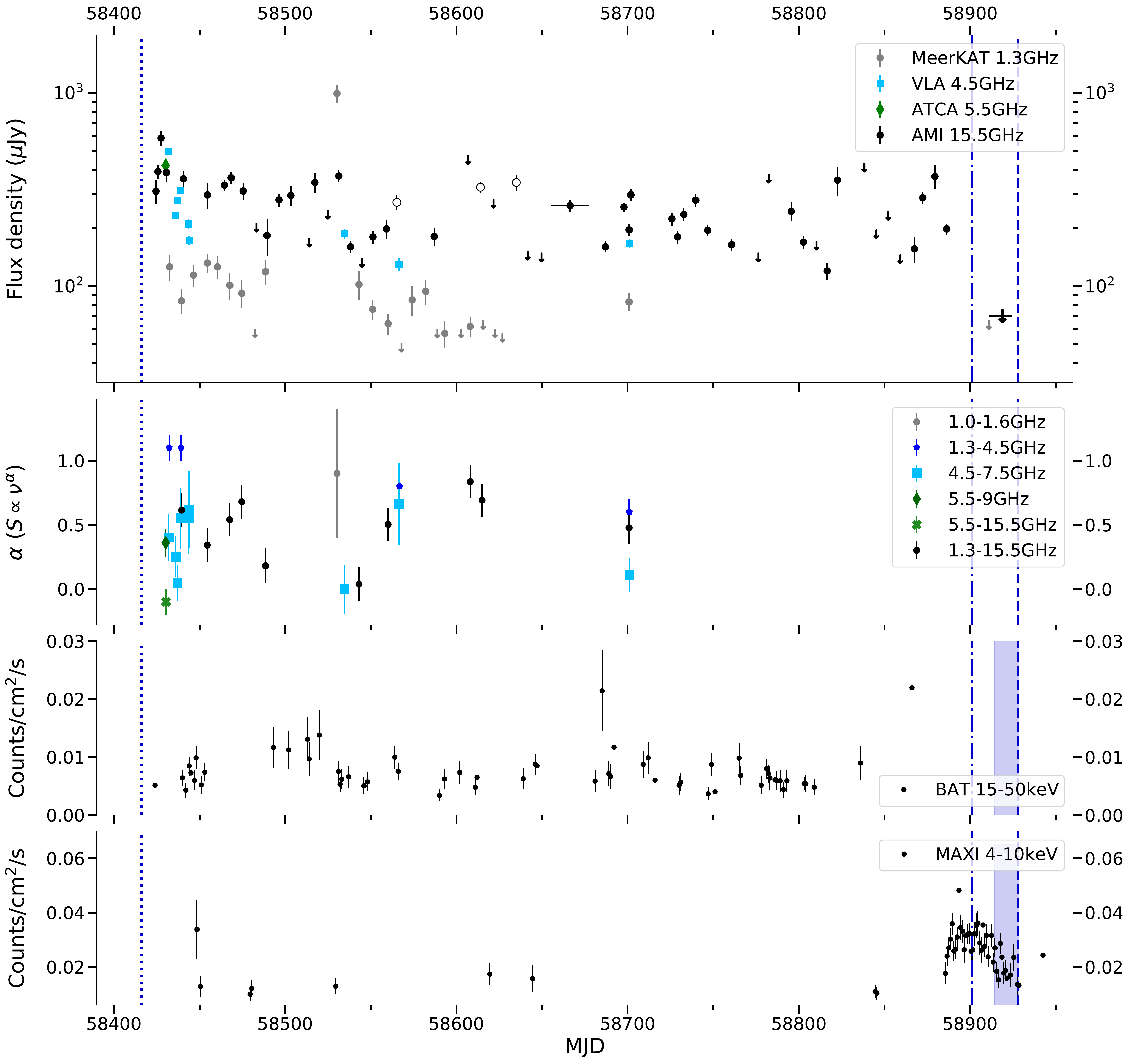}
    \caption{The upper panel of this plot shows the light curve for all AMI-LA and MeerKAT radio observations. MeerKAT detections are denoted by grey circles and AMI detections are black circles. There are three AMI detections denoted with black unfilled circles denoting epochs where uv-subtraction was performed to remove emission from a contaminating source as described in \ref{subsec:AMI}. Three sigma upper limits are marked with downwards facing triangles in black and grey for MeerKAT and AMI-LA respectively. The light blue squares and green diamonds are the VLA and ATCA data points from \citet{2020MNRAS.496.4127V}. The second panel shows the spectral indices for all quasi-simultaneous observations. Quasi-simultaneous observations are defined where the observations in the two bands are within 24 hours of each other. We only calculated spectral indices where there is a detection in at least one of the bands. The errors are calculated following \citet{2018MNRAS.473.4122E}. We see a flat/ inverted optically thick spectrum in all epochs. The lower two panels show the daily count rate from BAT (15-50\,keV) and MAXI/GSC(4-10\,keV), respectively. We only plot the data points where we obtain 3$\sigma$ detections. Despite MAXI/GSC monitoring the position of J1858 pre-outburst, no detection is made until the outburst begins. The MAXI/GSC light curve clearly shows a count rate increase around MJD 58880 indicates a state transition which was also seen by NICER \citep{2020ATel13536....1B}. The vertical dotted and dashed lines denotes the beginning and end of the outburst, respectively \citep{Krimm2018, 2020ATel13719....1S}. The dotted-dashed line indicates the time at which the state transition was observed by NICER on MJD 58901 \citep{2020ATel13536....1B}. The shaded region is the time over which Type I X-ray bursts were detected \citep{2020MNRAS.499..793B}. Figures \ref{fig:early} and \ref{fig:flare_zoom} show the periods MJD 58420 to MJD 58447 and MJD 58510 to 58550 for clarity.}
    \label{fig:lc}
\end{figure*}

To monitor the soft X-ray emission from J1858, we used the data between 4 and 10\,keV from the Monitor of All-sky X-ray Image mission \citep[MAXI/GSC,][]{2009PASJ...61..999M}. We use the daily count rate shown in the bottom panel of Figure \ref{fig:lc} and \ref{fig:flare_zoom}. 
 
To compare the radio and X-ray luminosities of J1858 during the `soft' and `hard' states, we also converted the average 4-10\,keV count rate to a 1-10\,keV luminosity for the period where we see a significant increase in soft X-ray flux using a black body temperature of 2.0\,keV \citep{2020ATel13536....1B}. The same N\textsubscript{H} value that was used to convert the BAT count rates was used here to calculate the unabsorbed fluxes as well. Both the BAT and MAXI/GSC converted luminosities are used in Figure \ref{fig:lr-lx}. We note that both the MAXI/GSC and BAT luminosities are subject to increased uncertainties due to variable N\textsubscript{H} measured throughout the outburst.

\section{Results}
\label{section:results}

\subsection{Radio Light Curves}
\label{subsec:radio_lc}

\begin{figure}
    \centering
    \includegraphics[width=\columnwidth]{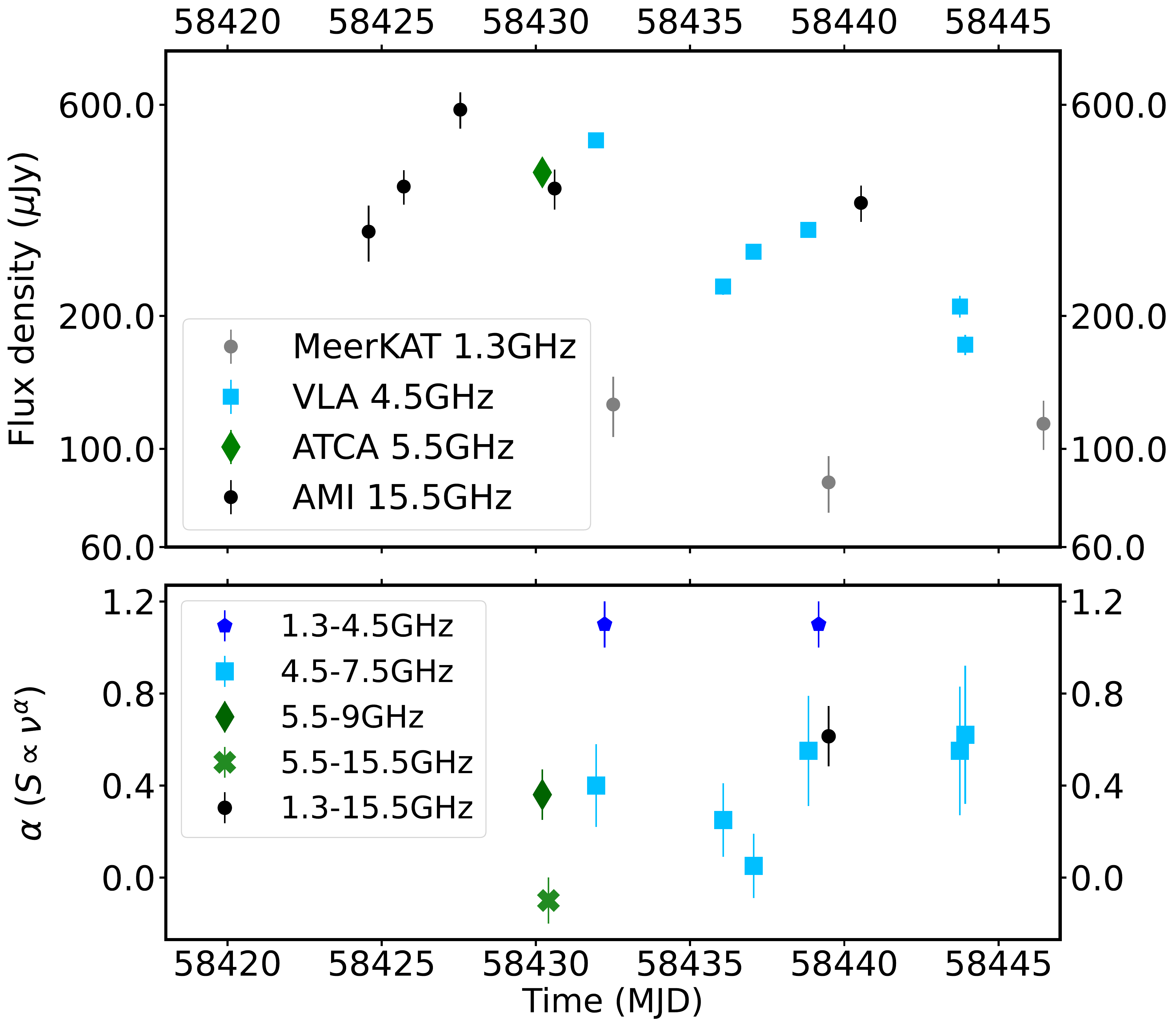}
    \caption{Subset of the top two panels of Figure \ref{fig:lc} between MJD 58420 and 58447 shown to demonstrate, more clearly, the epoch to epoch radio variability observed early on in the outburst. }
    \label{fig:early}
\end{figure}

The upper panel of Figure \ref{fig:lc} shows the MeerKAT 1.3\,GHz (grey circles) and AMI-LA 15.5\,GHz (black circles) light curves from MJD 58424 to 58924 (November 2018 to March 2020). The errorbars associated with the AMI-LA and MeerKAT data points are calculated by adding the fitting error and calibration error in quadrature. We use a 5 and 10 per cent calibration error for AMI-LA and MeerKAT, respectively. Three sigma upper limits are indicated with downwards facing arrows in the same colours as the detections for each telescope. We have also added VLA (4.5\,GHz) and ATCA (5.5\,GHz) data points from \citet{2020MNRAS.496.4127V}, shown as blue squares and green diamonds, respectively. We also show the radio light curve and spectral indices between MJD 58420 and 58450 in Figure \ref{fig:early}, where the radio coverage is denser earlier on in the outburst.

The first AMI-LA observation took place eight days after the outburst started \citep[blue dashed line in Figure \ref{fig:lc} and \ref{fig:early},][]{Krimm2018}. The AMI-LA light curve shows an initial flare that reached a peak flux density of almost 600$\mu$Jy at MJD 58427, seen more clearly in Figure \ref{fig:early}. We define a flare as a sudden, short period of increased radio flux density compared to the average level measured throughout the outburst. At the peak of the flare, we searched for shorter term variability by splitting the epoch into four one-hour long sections. J1858 showed $\sim$25\% variability on hour long timescales spanning between 450 and 800$\mu$Jy. Due to sensitivity constraints, we were unable to search for variability on timescales of less than one hour, like that seen in \citet{2020MNRAS.496.4127V}. For the rest of the outburst, the radio emission at 15.5\,GHz is seen to be variable on a timescale of weeks with a series of detections and non-detections all around 300$\mu$Jy.

Due to the delayed start time of the monitoring program with MeerKAT, we were unable to observe the initial flare at 1.3\,GHz. Like AMI-LA, the MeerKAT light curve shows slow flux variability on the week-time-scales, but at a lower flux level, around 100$\mu$Jy, except for a second flare observed at MJD 58530, which reached 1mJy at 1.3\,GHz. It is likely we missed the peak of the flare with MeerKAT given that we only observed J1858 for 15 minutes once a week, as a result the peak could have been much higher. We divided the MeerKAT epoch into shorter (five 3\,minute) sections in order to search for short timescale variability, we found the flux density to be constant across the epoch. During the period of the MeerKAT flare, there is no change in flux density the AMI-LA light curve. The flare seen with MeerKAT is discussed further in section \ref{subsubsec:flare}. 

\begin{figure}
    \centering
    \includegraphics[width=\columnwidth]{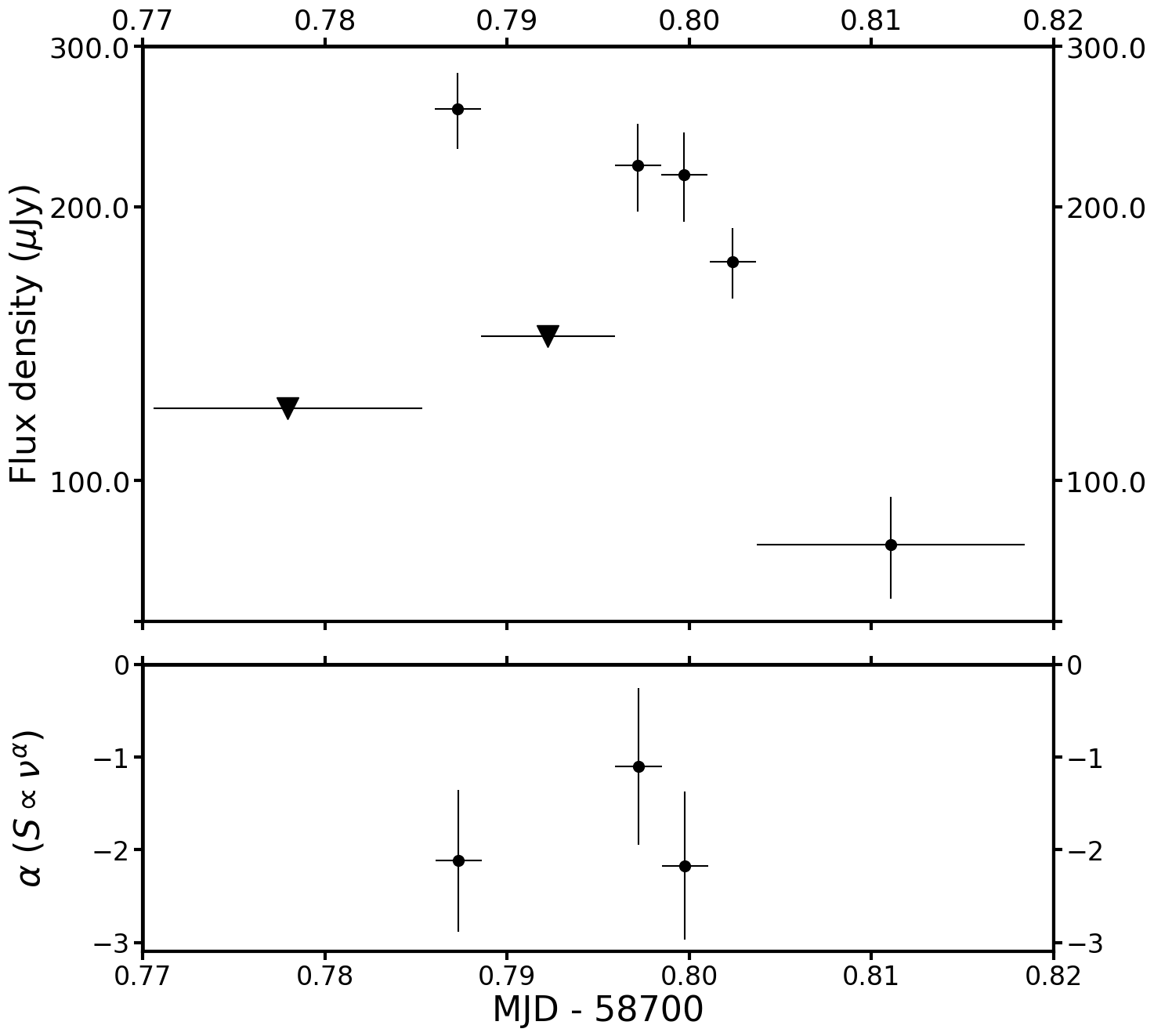}
    \caption{Data from an hour long observation with MeerKAT made as part of a multi-wavelength campaign including the Hubble Space Telescope (Castro-Segura et al, \textit{in press}). The top panel shows the 1.3\,GHz light curve for the observation. We observe variability on timescales of minutes, similar to that seen by \citet{2020MNRAS.496.4127V}. The bottom panel shows the in-band spectral index for the sub-integrations where the J1858 was bright enough to detect in both halves of the band. In both panels the horizontal error bars correspond to the duration of the integration. The vertical errorbars are the same as in Figures \ref{fig:lc}, \ref{fig:early} and \ref{fig:flare_zoom}.}
    \label{fig:MeerKAT}
\end{figure}

Following the second flare, the flux densities measured at 1.3\,GHz are fainter than at the beginning of the outburst and are all below 100$\mu$Jy. From MJD 58588, we obtained more 3$\sigma$ upper limits than detections and so on MJD 58626, weekly monitoring of J1858 with MeerKAT stopped. We later obtained a single hour long epoch, MJD 58700, as part of a multi-wavelength observing campaign. The signal to noise in the observation was high enough signal to split the observation into shorter integrations to search for short term variability. The resulting light curve is shown in the top panel of Figure \ref{fig:MeerKAT}. During the 60 minute observation, we detected two flares, the first lasting only two minutes, the second, about 30 minutes.

Our final radio detection was made on MJD 58886 with AMI-LA, around the same time as observations made with NICER and MAXI/GSC reported an increase in soft X-ray flux (see the bottom panel of Figure \ref{fig:lc}) and a state transition \citep{2020ATel13536....1B}. After the transition, we only measure 3$\sigma$ upper limits at the same flux level or higher than when J1858 was in the hard state. We have concatenated the final three AMI-LA upper limits, to place deeper constraints on any soft state jet quenching with respect to the hard state. Combining the last three data points gives a 3$\sigma$ upper limit of 70$\mu$Jy. We do not show the results of similar concatenation on either the hard state MeerKAT or AMI-LA epochs where we obtain a number of successive non-detections, in either case concatenating the data set did not further reduce the noise in the field.  

By MJD 58928 (20 March 2020), optical monitoring of J1858 suggested that the source was entering quiescence (blue dashed line in Figure \ref{fig:lc}) and was further confirmed by multiple \textit{Swift}-X-ray Telescope (XRT) observations showing continued low count rates \citep{2020ATel13719....1S, 2020ATel13725....1P}.

\subsection{Radio Spectrum}
The second panel of Figure \ref{fig:lc} shows radio spectral index measurements, where the spectral index $\alpha$ is defined as Flux density S$_{\nu} \propto \nu^{\alpha}$ ($\nu$ is the observing frequency, see lower panel of Figure \ref{fig:early} for spectral index measurements between MJD 58420 and 58450). The errors (1$\sigma$) were calculated using the following:
$$ err(\alpha) = \frac{1}{\textrm{ln}(\nu_{1}/\nu_{2})}\sqrt{\left(\frac{err(F_1)}{F_1}\right)^2 + \left(\frac{err(F_2)}{F_2}\right)^2}$$ where $\alpha$ is the spectral index, $\nu_{1,2}$, $F_{1,2}$ and $err(F_{1,2})$ are the observing frequency, flux density and associated uncertainty in the lower(1) and upper (2) halves of the band, respectively \citep{2018MNRAS.473.4122E}. We calculate spectral indices where the two observations in different bands are made within 24\,hours of each other and we detect the source in at least one of the bands. The black circles are the spectral index measurements between MeerKAT and AMI-LA (1.3-15.5\,GHz), the blue squares are for MeerKAT-VLA (1.3-4.5\,GHz) and the dark green diamonds are for ATCA-AMI-LA (5.5-15.5\,GHz). 

\citet{2020MNRAS.496.4127V} published the measured radio flux densities in the two halves of the observing band for both the VLA (4.5-7.5\,GHz, turquoise hexagons) and ATCA (5.5-9\,GHz, green crosses), allowing for in-band spectral index measurements. We calculate the 1.0-1.6\,GHz MeerKAT for a single MeerKAT epoch (MJD 58530) where the source reached 1\,mJy (light blue circle). All calculations show the radio emission from J1858 to be consistent with being either inverted or flat, i.e. $\alpha \geq 0$.

The only time we obtained spectral index measurements where $\alpha\leq0$ was on MJD 58700. The bottom panel of Figure \ref{fig:MeerKAT} shows the in-band spectral indices for the MeerKAT observation on MJD 58700 for the periods where J1858 was bright enough in both halves of the band. The negative spectral index measurements were obtained only on timescales of less than 15 minutes. \citet{2020MNRAS.496.4127V} also measured variable spectral indices on short timescales with rapid changes with $\alpha\leq0$ and $\alpha\geq0$.

\subsection{X-ray Light Curve}
The bottom two panels of Figure \ref{fig:lc} show the daily average count rate from \textit{Swift}-BAT, between 15 and 50\,keV, and MAXI/GSC, between 4 and 10\,keV, respectively, for all detections that are 3$\sigma$ or greater.

The BAT X-ray light curve shows repeated, low-level, hard X-ray detections from the beginning of the outburst, MJD 58422 until 58800. When combined with quasi-persistent, self-absorbed radio detections, the detections from  BAT imply that J1858 is in the hard state. There are two periods during the outburst: MJD 58470 to 58520, and 58830 to $\sim$58880, where J1858 was close to the Sun. During this period, BAT and MAXI/GSC, which scan large portions of the sky, were still able to observe but the resulting data points have larger associated uncertainties. Our final BAT detection is on MJD 58866.

From MJD 58880 onwards, after the final BAT detection, we began to obtain regular detections with MAXI/GSC. The bottom panel of Figure \ref{fig:lc} shows a sharp increase in soft X-ray flux (bottom panel of Figure \ref{fig:lc}) that lasts until around MJD 58940. The observed flux increase is consistent with the NICER and MAXI observations made on MJD 58897 reporting a change in X-ray spectral state \citep{2020ATel13536....1B, 2020ATel13455....1N}. After reaching a peak around MJD 58900, the MAXI/GSC light curve shows a gradual decay as J1858 returns to quiescence \citep[the blue dashed line in Figure \ref{fig:lc}][]{2020ATel13719....1S}. Previous to the state transition, we observed regular BAT detections indicative of J1858 being in the hard state. Between the soft state and quiescence, there are no further hard X-ray detections. However, we cannot conclusively say that J1858 did not return to the hard state. When X-ray binaries return to the hard state before quiescence, it may be at a lower flux level than when the system first enters the outburst \citep{2004ARA&A..42..317F}. It is possible that the hard state emission was too faint for BAT to detect.

\section{Discussion}
\label{sec:discussion}

We have monitored the newly discovered neutron star XRB J1858 throughout its outburst, lasting from MJD 58242 to 58924 (late 2018 to early 2022). Our radio observations show faint emission consistent with a quasi-persistent (but variable), self-absorbed jet. X-ray monitoring with both BAT and MAXI/GSC show little evidence of strong variability on day-long timescales. On two separate occasions we observe flares, one with AMI at the beginning of the outburst and a second with MeerKAT around MJD 58530. Here, we place our observations in the context of the wider XRB population. 

\subsection{Jet emission}

We detected J1858 in approximately 60\% of our observations with AMI-LA, with flux densities in a range between 150 and 350$\mu$Jy (with the exception of the flare around MJD 58430). We note that several non-detections obtained with AMI-LA are not particularly constraining given they correspond to flux densities higher than in the narrow range over which J1858 is detected. Combined with the positive spectral index measurements (with the exception of MJD 58700), the radio detections point towards radio emission produced by a compact, self-absorbed jet \citep{2003MNRAS.342L..67M}. We are only able to search for short (minute) timescale variability, like that found by \citet{2020MNRAS.496.4127V}, in a single epoch: MJD 58700, see Figure \ref{fig:MeerKAT}. Some of the VLA and ATCA epochs overlap with our observations. Those observations show short term variability whereas our observations do not, mainly due to lack of signal to noise. The short term variability is accompanied by changing, often negative, spectral indices \citep{2020MNRAS.496.4127V}. It may be the case that for most of the outburst where our observations average over periods of 15 minutes or longer, any short time-scale variability and change in spectral index may average out as a relatively steady and self-absorbed jet.

Similar radio properties have been observed in V404 Cygni, a black hole XRB that J1858 has been compared to previously due to similar X-ray signatures \citep{2018ATel12158....1L}. In V404 Cygni, the time averaged spectrum showed a self-absorbed component like what we see in the second panel of Figure \ref{fig:lc}, but on shorter timescales (minutes) there are transitions from optically thick to thin like seen in Figure \ref{fig:MeerKAT}. In V404 Cygni, it was suggested that the rapid spectral variability arose from intrinsic fluctuations in the accretion flow \citep{2016ApJ...821..103R}.

Very few other neutron star XRB outbursts have had comparable, long term, radio observing campaigns with such high signal to noise ratios so that the short timescale spectral index and flux density variability can be studied. Observations of neutron star XRB Scorpius X-I, a Z source, by \citet{2001ApJ...558..283F} showed $\sim$hour timescale variability in both flux density and spectral index. This is perhaps the most similar behaviour to that observation in J1858. Regular radio observations of GRS 1747-312 showed strong swings in spectral index, on longer timescales: over a period of three months \citep{2021ApJ...923...88P}. One the other hand, XTE J1701-462 was observed every three days for three months with ATCA in 2006 and showed a steady flat/ inverted spectral index, similar what has been observed in J1858 \citep{2007MNRAS.380L..25F}.

Persistent radio jets and power law components in the X-ray spectra are clear signatures of hard state XRBs. As mentioned in Section \ref{sub:J1858}, observations of J1858 at X-ray energies shows that the source does not appear to have a canonical hard X-ray state, instead, the X-ray emission is highly variable with hard emission interspersed with soft flares \citep{2020ApJ...890...57H}. The quasi-persistent, self-absorbed radio emission observed on long timescales is perhaps the best indication that J1858 is in the hard state.

Our final radio detection is on MJD 58886. Around the same time, our BAT light curve shows no further detections, instead we see a significant increase in flux in the MAXI 4-10\,keV light curve, as shown in the lower two panels of Figure \ref{fig:lc}. Such a change indicates that J1858 transitioned to a state that is dominated by soft X-ray photons. Similarly to when in the hard state, \citet{2020ATel13536....1B}'s NICER observations between 0.6 and 12\,keV demonstrates that J1858 also does not show a canonical soft state, however, the spectrum can no longer be described using a single power law, a soft component (from a disc) is also required, thus indicating a potential state change.

The final radio detection with AMI-LA on MJD 58886 is fainter by a factor of two compared to the detection a week earlier, hinting at a possible jet quenching during the transition to the softer state. We concatenate the final three AMI-LA non-detections and obtain a much deeper 3$\sigma$ upper limit of 70$\mu$Jy. In comparison to the average AMI-LA flux density measured whilst J1858 was in the hard state, the jet emission from J1858 was quenched by at least a factor of four upon entering the softer state, if the reduction in observed radio flux density is due to quenching from a state transition. Evidence of jet quenching has been observed in some other neutron star XRBs, but there are also cases where bright radio emission has been detected in the soft state indicating that jet quenching is not a global phenomena in neutron star systems \citep{2004MNRAS.351..186M, 2017MNRAS.470.1871G}. Further observations of neutron star systems in both hard and soft states are required to better understand any jet quenching.

\begin{figure}
    \centering
    \includegraphics[width = \columnwidth]{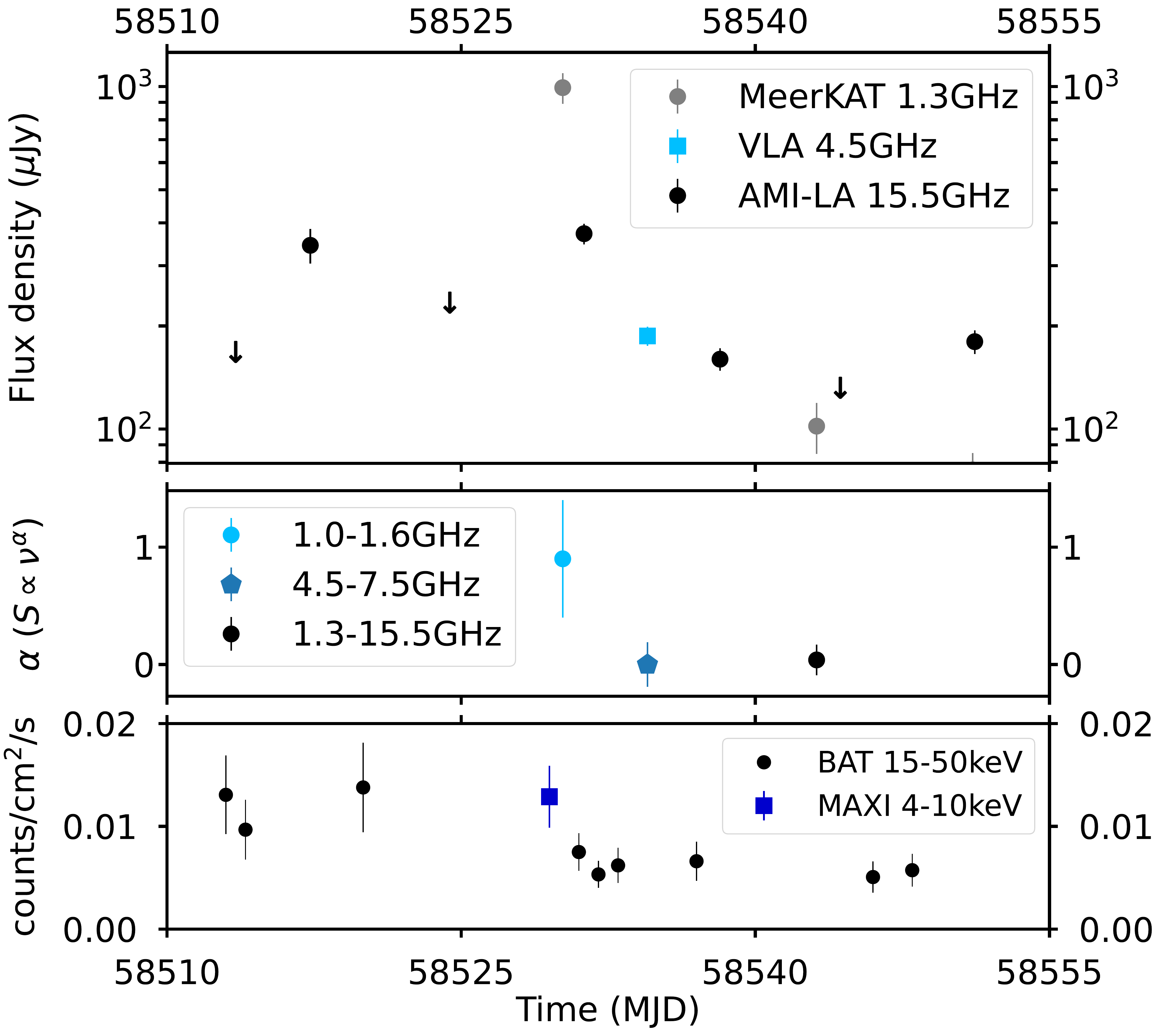}
    \caption{A subsection of Figure \ref{fig:lc} showing the radio and X-ray light curves for 20 days either side of the 1\,mJy flare observed with MeerKAT. Combining observations from 3 different observing frequencies we see a a sharp rise and a decay. Also shown are the radio spectral index measurements for the same period. The X-ray light curve remains fairly consistent during the period of the flare. }
    \label{fig:flare_zoom}
\end{figure}

\subsubsection{Optically-thick flare}
\label{subsubsec:flare}

The radio emission associated with J1858 has been fairly constant on long (weeks-months) timescales at faint flux ($\sim$100\,$\mu$Jy/beam) density level, similar to other self-absorbed compact jets observed in other neutron star XRBs. During J1858's outburst, there were two occasions where the radio emission deviated from this pattern in the form of optically thick flares (MJD 58430 and 58530). Radio flares are usually associated with state transitions in both black hole and some neutron star XRBs, tracking the evolution of the flares show an transition from optically thick to thin synchrotron emission \citep{1994Natur.371...46M, 2004Natur.427..222F, 2001ApJ...558..283F}. During the transition, flares are thought to occur when discrete blobs of synchrotron emitting plasma are ejected from the accreting compact object. Depending on the how quickly the transition occurred, we may also expect to see a drop in the hard X-ray flux, due to the transition to a softer X-ray spectral state. Figure \ref{fig:flare_zoom} shows the radio and X-ray light curves, as well as the radio spectral index measurements for the period 20 days either side of the optically thick flare observed with MeerKAT. There is a single soft X-ray detection during the 40 days (bottom panel of Figure \ref{fig:flare_zoom}), just before we detect the flare with MeerKAT. BAT detects J1858 repeatedly during this period (MJD 58510 - 58550), with a BAT count rate that is approximately the same either side of the flare (also see panel three of Figure \ref{fig:lc}). It may be likely that a fast transition occurred and could have easily been missed given how faint J1858 is.

On MJD 58530, the flux density of J1858 observed with MeerKAT reached 1.0$\pm$0.1\,mJy, about an order of magnitude brighter than the average MeerKAT flux density for the rest of the outburst. The 1.0-1.6\,GHz spectral index of the flare was $\alpha = 0.9\pm0.5$ (consistent with optically thick emission, see middle panel of Figure \ref{fig:flare_zoom}). By observing an optically thick flare, we can place lower limits on the size of ejecta at the peak of the flare, infer the minimum energy, magnetic field associated with the minimum energy and the brightness temperature \citep{Fender2019}. At a distance of 12.8$^{+0.7}_{-0.6}$\,kpc, \citep{2020MNRAS.499..793B}, a peak flux of 0.99\,mJy at a frequency of 1.3\,GHz corresponds to a brightness temperature of $5\times10^{10}$K and energy of $5\times10^{37}$\,erg. The magnetic field associated associated with the minimum energy is 0.2\,G. At the peak of the flare, the emission region size is 10\textsuperscript{13}\,cm. Our results are at a similar order of magnitude or slightly lower than those calculated for black hole XRBs V404 Cygni, Cygnus X$-$3, GRS 1915$+$105 and MAXI J1631$-$472 \citep{Fender2019, 2021MNRAS.501.5776M}. We obtained a minimum energy four orders of magnitude higher compared to that observed in cataclysmic variable SS Cygni \citep{2019MNRAS.490L..76F}. Placing J1858 in the context of other transient compact systems: the radio flares observed in black hole and neutron star XRBs correspond to ejections of similar sizes and energies, independent of the nature of the compact object. Cataclysmic variables, on the other hand, launch much less powerful ejections. We note that this comparison is based on a limited sample size.  

Despite observing the radio flare, we observe no evidence of movement of the radio source associated with J1858 in the MeerKAT images subsequent to the 1\,mJy flare. Such motion would imply the ejection of plasma of a similar energy and duration as that observed in multiple black hole systems by MeerKAT \citep{2020NatAs...4..697B, 2021MNRAS.504..444C}. However, given the large distance to J1858, combined with the large MeerKAT beam size ($\sim7$"), it is not surprising that we do not see any movement of the radio source in the MeerKAT images. In order to resolve any proper motion, we would require VLBI. In the AMI-LA and MeerKAT observations made the following week, the optically thick jet feature had returned with a flux level $\sim 100\mu$Jy fainter than before the flare. The return of the compact jet component indicates that any transition to the soft state occurred on a timescale of days.

\subsection{Radio X-Ray Correlation}

\begin{figure*}
    \centering
    \includegraphics[width = \textwidth]{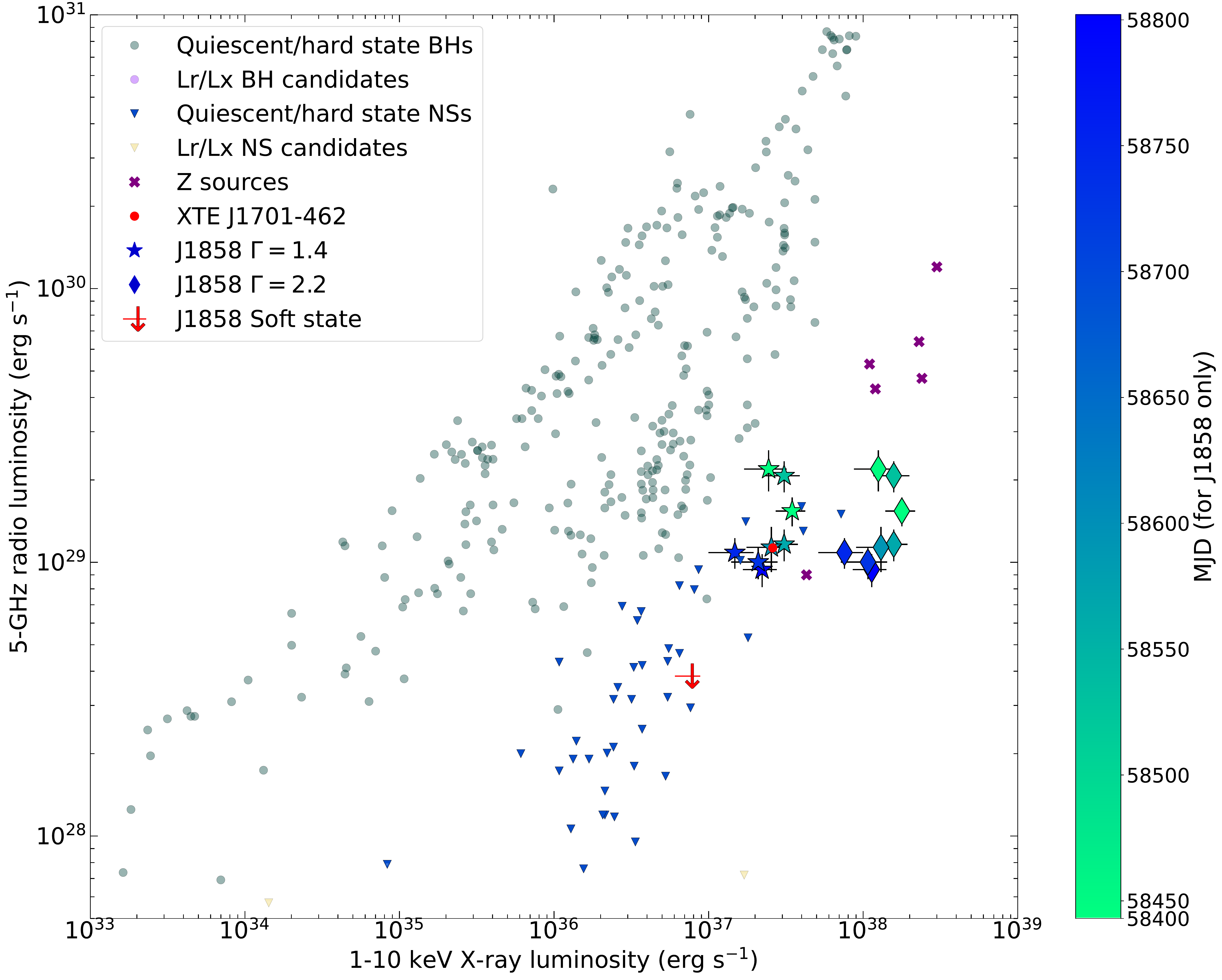}

    \caption{Radio/X-ray luminosity plane for a range of black hole and neutron star XRBs in the hard state (or equivalent) \citep{arash_bahramian_2018_1252036}. We have also added a sample of Z sources from \citet{2006MNRAS.366...79M}, shown with purple crosses as well as XTE J1701$-$462, a neutron star XRB that has shown both Atoll and Z source phases \citep[red circle][]{2007MNRAS.380L..25F}. Using a distance of 12.8$^{+0.7}_{-0.6}$\,kpc, we calculated the 5\,GHz and 1-10keV luminosities for every quasi-simultaneous observation between \textit{Swift}-BAT and MeerKAT, VLA or AMI-LA \citep{2020MNRAS.499..793B}. We have calculated the 5\,GHz luminosity by assuming a spectral index of $\nu^{0.5}$, the average spectral index calculated from our observing campaign. The X-ray luminosities are calculated using two different photon indices, 1.4 and 2.2 (diamonds and stars, respectively), to reflect the range of values measured throughout the outburst \citep{Bozzo2018}. The errorbars on the X-ray luminosities consider the count rate and distance uncertainties, we do not consider the uncertainty induced by the variable absorption because in the \textit{Swift}-BAT observing band (15-50keV, where the X-ray data points were obtained), the variable N\textsubscript{H} affects are negligible. The colour gradient shows the movement of J1858 on the radio/X-ray plane as the outburst evolved. Three AMI-LA observations took place after J1858 transitions in to a softer state. We have concatenated these observations and combined with MAXI/GSC observations made in the same period to produce the downwards facing red arrow.}
    \label{fig:lr-lx}
\end{figure*}

Using the detections of self-absorbed radio emission as an indication that J1858 is in the `hard state', we can combine these detections with hard X-ray detections and compare J1858 to other hard state XRBs. Figure \ref{fig:lr-lx} shows the 5\,GHz radio and 1-10keV X-ray luminosities for a compilation of hard state neutron star (downwards facing blue triangles) and black hole XRBs \citep[black circles,][]{arash_bahramian_2018_1252036}. We have also plotted a selection of Z sources from \citet[][purple crosses]{2006MNRAS.366...79M}. The red circle denotes the radio and X-ray luminosities of XTE J1701$-$462, an interesting neutron star system which has shown both Atoll and Z-source-like phases \citep{2007MNRAS.380L..25F}. Overlaid are the radio and X-ray luminosities for every epoch where we have quasi-simultaneous radio and BAT detections.

The very high and varying column density in J1858, \citep{2020ApJ...890...57H}, makes estimating the intrinsic X-ray luminosity difficult. \citet{2020MNRAS.496.4127V} showed that the XRT fluxes under-represent the intrinsic emission by a factor of 2-4 due to significant absorption along the line of sight. The use of NuSTAR, which observes at a higher energy range than XRT allowed for the calculation of the ionizing X-ray luminosity at 2$\times$10\textsuperscript{38}erg\,s\textsuperscript{-1}, \citep{2020MNRAS.499..793B}, which is approximately 30 times higher than the luminosities derived from the XRT observations. This shows that absorption cannot fully explain the flux discrepancy found between BAT and XRT. A large fraction of the emission is scattered out of our line of sight, which may be due to the high inclination of the source \citep[>70$^{\circ}$,][]{2021MNRAS.503.5600B}. We find that the BAT fluxes are less affected by the significant absorption and scattering, the luminosities we obtain are similar to that obtained by \citet{2020MNRAS.499..793B}. Therefore, in Figure \ref{fig:lr-lx}, we extrapolate the BAT fluxes measured between 15-50\,keV to 1-10keV using two different photon indices to reflect the varying spectrum that has been reported: 1.4 and 2.2, shown as diamonds and stars, respectively.

To calculate the radio luminosity, we scale the radio detection to 5\,GHz using a spectral index of 0.5, the average spectral index measured throughout the outburst. We note that a flat spectral index is usually assumed when scaling radio luminosities to 5\,GHz both to consider the difference in radio spectral index measurements on the radio loud and quiet black hole branches and when a spectral index is not available \citep{2018MNRAS.473.4122E}. If we were to use a flat spectral index, our MeerKAT luminosities would be about a factor of two lower and the luminosities from our AMI-LA data points would about a factor of two higher.

The colour gradient in Figure \ref{fig:lr-lx} shows how J1858 moves across the radio/X-ray correlation throughout the outburst. For a given photon index, the source follows an approximately semi-circular path on the radio/X-ray plane, becoming less radio-luminous by about a factor of three as the outburst progresses. Unlike the radio luminosity, the 1-10\,keV luminosity remains fairly constant. However, this is all under the assumption that the X-ray photon index is constant across all epochs. It is most likely that the X-ray luminosity and photon index varies significantly throughout the outburst. We can use J1858 to sample potentially new luminosity parameter space between the hard state neutron stars and Z sources in Figure \ref{fig:lr-lx}.

J1858 sits between the atoll and Z sources in terms of both X-ray and radio luminosities in Figure \ref{fig:lr-lx}. It is at the low luminosity end of the Z source distribution but also at the high luminosity end of the atoll group. The long term radio light curve shows emission consistent with a compact self-absorbed jet, similar to those observed in other atoll sources. However, on short timescales, we observe some rapid flaring (Figure \ref{fig:MeerKAT}), which is Z source-like behaviour, the flaring is only observed for a short period and we cannot be certain that such behaviour is present through the outburst. We note that if one assumes $\Gamma = 1.4$, J1858 also overlaps perfectly with the position of XTE J1701-462. XTE J1701-462 is a neutron star XRB that was initially classified as a Z source \citep{2006ATel..725....1H}. Subsequent analysis of RXTE observations that covered the duration of the outburst showed that XTE J1701-462 evolved from a Z source to an atoll source \citep{2007ApJ...656..420H}. Radio observations showed a fairly constant (few 100$\mu$Jy) point source with a inverted spectral index in almost all epochs except when at the highest accretion rate, similar to what has been observed from J1858 \citep{2007MNRAS.380L..25F}. Therefore, despite the wealth of observations and data, we are unable to determine whether J1858 is an atoll or Z source. 


We have also placed our deep radio upper limit on Figure \ref{fig:lr-lx} corresponding to the time after the spectral state change \citep[around MJD 58900,][]{2020ATel13536....1B}. The resulting data point is shown as a red downwards facing arrow. As J1858 moves into the soft-state, the radio luminosity continues to drop reaching an 3$\sigma$ upper limit of $\sim$4$\times$10\textsuperscript{28}erg\,s\textsuperscript{-1}. The radio luminosity in the soft-state is about a factor of four fainter than the average flux density measured with AMI-LA throughout the outburst. The X-ray luminosity also decreases by at least an order of magnitude ($\Gamma$-dependent).

\section{Conclusions}

We have presented near weekly radio observations of the outburst of neutron star X-ray binary J1858.6-0815 that started in October 2018. Our radio observations show self-absorbed emission, consistent with a quasi-steady compact jet as expected in the hard X-ray spectral state, despite J1858 not appearing to be in a canonical hard state. The radio light curves show little significant long term variability, with the exception of MJD 58427 and 58530, where we see two flares. During these periods, we see no clear X-ray link to the radio flares. On shorter timescales, MJD 58700, there is significant variability in terms of flux density and spectral index \citep[also shown in ][]{2020MNRAS.496.4127V}. We compare J1858 to other XRBs on the radio/X-ray correlation. At a distance of 12.8$^{+0.7}_{-0.6}$kpc, J1858 appears to be a very radio luminous atoll-type or faint Z source neutron star binary yet placed in the radio/X-ray plane, and therefore of fundamental importance in establishing the slope and normalisation of the correlation for neutron stars. This in turn allows us to probe ever deeper the differences between accretion and jet production in neutron stars and blacks holes.


\section*{Acknowledgements}

We thank the referees for their useful comments. The MeerKAT telescope is operated by the South African Radio Astronomy Observatory, which is a facility of the National Research Foundation, an agency of the Department of Science and Innovation. We acknowledge the use of public data from the Swift data archive. This research has made use of MAXI data provided by RIKEN, JAXA and the MAXI team. L. Rhodes acknowledges the support given by the Science and Technology Facilities Council through an STFC studentship. J.v.d.E. is supported by a Lee Hysan Junior Research Fellowship awarded by St. Hilda's College. This work was supported by the Oxford Centre for Astrophysical Surveys, which is funded through generous support from the Hintze Family Charitable Foundation. GRS acknowledge support from Natural Sciences and Engineering Research Council of Canada (NSERC) Discovery Grants RGPIN-2016-06569 and RGPIN-2021-04001.

\section*{Data Availability}

The radio data discussed in this paper are available in the Appendix. The data that forms the \textit{Swift}-BAT light curve is available from the \textit{Swift}-BAT Hard X-ray Transient Monitor repository\footnote{https://swift.gsfc.nasa.gov/results/transients/}. The MAXI/GSC data are provided by RIKEN, JAXA and the MAXI team\footnote{http://maxi.riken.jp/star\_data/J1858$-$082/J1858$-$082.html}.

\bibliographystyle{mnras}
\bibliography{bibliography.bib} 

\begin{thebibliography}{}
\makeatletter
\relax
\def\mn@urlcharsother{\let\do\@makeother \do\$\do\&\do\#\do\^\do\_\do\%\do\~}
\def\mn@doi{\begingroup\mn@urlcharsother \@ifnextchar [ {\mn@doi@}
  {\mn@doi@[]}}
\def\mn@doi@[#1]#2{\def\@tempa{#1}\ifx\@tempa\@empty \href
  {http://dx.doi.org/#2} {doi:#2}\else \href {http://dx.doi.org/#2} {#1}\fi
  \endgroup}
\def\mn@eprint#1#2{\mn@eprint@#1:#2::\@nil}
\def\mn@eprint@arXiv#1{\href {http://arxiv.org/abs/#1} {{\tt arXiv:#1}}}
\def\mn@eprint@dblp#1{\href {http://dblp.uni-trier.de/rec/bibtex/#1.xml}
  {dblp:#1}}
\def\mn@eprint@#1:#2:#3:#4\@nil{\def\@tempa {#1}\def\@tempb {#2}\def\@tempc
  {#3}\ifx \@tempc \@empty \let \@tempc \@tempb \let \@tempb \@tempa \fi \ifx
  \@tempb \@empty \def\@tempb {arXiv}\fi \@ifundefined
  {mn@eprint@\@tempb}{\@tempb:\@tempc}{\expandafter \expandafter \csname
  mn@eprint@\@tempb\endcsname \expandafter{\@tempc}}}

\bibitem[\protect\citeauthoryear{Bahramian et~al.,}{Bahramian
  et~al.}{2018}]{arash_bahramian_2018_1252036}
Bahramian A.,  et~al., 2018, {Radio/X-ray correlation database for X-ray
  binaries}, \mn@doi{10.5281/zenodo.1252036}, \url
  {https://doi.org/10.5281/zenodo.1252036}

\bibitem[\protect\citeauthoryear{Belloni \& Motta}{Belloni \&
  Motta}{2016}]{2016ASSL..440...61B}
Belloni T.~M.,  Motta S.~E.,  2016, Transient Black Hole Binaries.
Springer Berlin Heidelberg, Berlin, Heidelberg, pp 61--97,
  \mn@doi{10.1007/978-3-662-52859-4_2}, \url
  {https://doi.org/10.1007/978-3-662-52859-4_2}

\bibitem[\protect\citeauthoryear{{Bozzo}, {Ferrigno}, {Savchenko}, {Ducci}  \&
  {Kuulkers}}{{Bozzo} et~al.}{2018}]{Bozzo2018}
{Bozzo} E.,  {Ferrigno} C.,  {Savchenko} V.,  {Ducci} L.,   {Kuulkers} E.,
  2018, The Astronomer's Telegram, \href
  {https://ui.adsabs.harvard.edu/abs/2018ATel12167....1B} {12167}

\bibitem[\protect\citeauthoryear{{Bright}, {Fender}, {Motta}, {Rhodes},
  {Titterington}  \& {Perrott}}{{Bright} et~al.}{2018}]{2018ATel12184....1B}
{Bright} J.,  {Fender} R.,  {Motta} S.,  {Rhodes} L.,  {Titterington} D.,
  {Perrott} Y.,  2018, The Astronomer's Telegram, \href
  {https://ui.adsabs.harvard.edu/abs/2018ATel12184....1B} {12184, 1}

\bibitem[\protect\citeauthoryear{{Bright} et~al.,}{{Bright}
  et~al.}{2020}]{2020NatAs...4..697B}
{Bright} J.~S.,  et~al., 2020, \mn@doi [Nature Astronomy]
  {10.1038/s41550-020-1023-5}, \href
  {https://ui.adsabs.harvard.edu/abs/2020NatAs...4..697B} {4, 697}

\bibitem[\protect\citeauthoryear{{Buisson} et~al.,}{{Buisson}
  et~al.}{2020a}]{2020MNRAS.499..793B}
{Buisson} D.~J.~K.,  et~al., 2020a, \mn@doi [\mnras] {10.1093/mnras/staa2749},
  \href {https://ui.adsabs.harvard.edu/abs/2020MNRAS.499..793B} {499, 793}

\bibitem[\protect\citeauthoryear{{Buisson}, {Altamirano}, {Remillard},
  {Arzoumanian}, {Gendreau}, {Gandhi}  \& {Vincentelli}}{{Buisson}
  et~al.}{2020b}]{2020ATel13536....1B}
{Buisson} D. J.~K.,  {Altamirano} D.,  {Remillard} R.,  {Arzoumanian} Z.,
  {Gendreau} K.,  {Gandhi} P.,   {Vincentelli} F.,  2020b, The Astronomer's
  Telegram, \href {https://ui.adsabs.harvard.edu/abs/2020ATel13536....1B}
  {13536, 1}

\bibitem[\protect\citeauthoryear{{Buisson} et~al.,}{{Buisson}
  et~al.}{2020c}]{2020ATel13563....1B}
{Buisson} D.~J.~K.,  et~al., 2020c, The Astronomer's Telegram, \href
  {https://ui.adsabs.harvard.edu/abs/2020ATel13563....1B} {13563}

\bibitem[\protect\citeauthoryear{{Buisson} et~al.,}{{Buisson}
  et~al.}{2021}]{2021MNRAS.503.5600B}
{Buisson} D.~J.~K.,  et~al., 2021, \mn@doi [\mnras] {10.1093/mnras/stab863},
  \href {https://ui.adsabs.harvard.edu/abs/2021MNRAS.503.5600B} {503, 5600}

\bibitem[\protect\citeauthoryear{{Carotenuto} et~al.,}{{Carotenuto}
  et~al.}{2021}]{2021MNRAS.504..444C}
{Carotenuto} F.,  et~al., 2021, \mn@doi [\mnras] {10.1093/mnras/stab864}, \href
  {https://ui.adsabs.harvard.edu/abs/2021MNRAS.504..444C} {504, 444}

\bibitem[\protect\citeauthoryear{Castro~Segura et~al.,}{Castro~Segura
  et~al.}{2022}]{CastroSegura2022}
Castro~Segura N.,  et~al., 2022, \mn@doi [Nature] {10.1038/s41586-021-04324-2},
  603, 52

\bibitem[\protect\citeauthoryear{{Degenaar} et~al.,}{{Degenaar}
  et~al.}{2016}]{2016MNRAS.461.4049D}
{Degenaar} N.,  et~al., 2016, \mn@doi [\mnras] {10.1093/mnras/stw1593}, \href
  {https://ui.adsabs.harvard.edu/abs/2016MNRAS.461.4049D} {461, 4049}

\bibitem[\protect\citeauthoryear{{Done}, {Gierli{\'n}ski}  \& {Kubota}}{{Done}
  et~al.}{2007}]{2007A&ARv..15....1D}
{Done} C.,  {Gierli{\'n}ski} M.,   {Kubota} A.,  2007, \mn@doi [\aapr]
  {10.1007/s00159-007-0006-1}, \href
  {https://ui.adsabs.harvard.edu/abs/2007A&ARv..15....1D} {15, 1}

\bibitem[\protect\citeauthoryear{{Espinasse} \& {Fender}}{{Espinasse} \&
  {Fender}}{2018}]{2018MNRAS.473.4122E}
{Espinasse} M.,  {Fender} R.,  2018, \mn@doi [\mnras] {10.1093/mnras/stx2467},
  \href {https://ui.adsabs.harvard.edu/abs/2018MNRAS.473.4122E} {473, 4122}

\bibitem[\protect\citeauthoryear{{Fender}}{{Fender}}{2001}]{Fender2001}
{Fender} R.~P.,  2001, \mn@doi [\mnras] {10.1046/j.1365-8711.2001.04080.x},
  \href {https://ui.adsabs.harvard.edu/abs/2001MNRAS.322...31F} {322, 31}

\bibitem[\protect\citeauthoryear{{Fender} \& {Belloni}}{{Fender} \&
  {Belloni}}{2004}]{2004ARA&A..42..317F}
{Fender} R.,  {Belloni} T.,  2004, \mn@doi [\araa]
  {10.1146/annurev.astro.42.053102.134031}, \href
  {https://ui.adsabs.harvard.edu/abs/2004ARA&A..42..317F} {42, 317}

\bibitem[\protect\citeauthoryear{Fender \& Bright}{Fender \&
  Bright}{2019}]{Fender2019}
Fender R.,  Bright J.,  2019, \mn@doi [\mnras] {10.1093/mnras/stz2000}, 489,
  4836

\bibitem[\protect\citeauthoryear{{Fender} et~al.,}{{Fender}
  et~al.}{1999}]{1999ApJ...519L.165F}
{Fender} R.,  et~al., 1999, \mn@doi [\apjl] {10.1086/312128}, \href
  {https://ui.adsabs.harvard.edu/abs/1999ApJ...519L.165F} {519, L165}

\bibitem[\protect\citeauthoryear{Fender, Belloni  \& Gallo}{Fender
  et~al.}{2004a}]{Fender2004}
Fender R.~P.,  Belloni T.~M.,   Gallo E.,  2004a, \mn@doi [\mnras]
  {10.1111/j.1365-2966.2004.08384.x}, 355, 1105

\bibitem[\protect\citeauthoryear{{Fender}, {Wu}, {Johnston}, {Tzioumis},
  {Jonker}, {Spencer}  \& {van der Klis}}{{Fender}
  et~al.}{2004b}]{2004Natur.427..222F}
{Fender} R.,  {Wu} K.,  {Johnston} H.,  {Tzioumis} T.,  {Jonker} P.,  {Spencer}
  R.,   {van der Klis} M.,  2004b, \mn@doi [\nat] {10.1038/nature02137}, \href
  {https://ui.adsabs.harvard.edu/abs/2004Natur.427..222F} {427, 222}

\bibitem[\protect\citeauthoryear{{Fender}, {Dahlem}, {Homan}, {Corbel}, {Sault}
   \& {Belloni}}{{Fender} et~al.}{2007}]{2007MNRAS.380L..25F}
{Fender} R.~P.,  {Dahlem} M.,  {Homan} J.,  {Corbel} S.,  {Sault} R.,
  {Belloni} T.~M.,  2007, \mn@doi [\mnras] {10.1111/j.1745-3933.2007.00350.x},
  \href {https://ui.adsabs.harvard.edu/abs/2007MNRAS.380L..25F} {380, L25}

\bibitem[\protect\citeauthoryear{{Fender}, {Homan}  \& {Belloni}}{{Fender}
  et~al.}{2009}]{2009MNRAS.396.1370F}
{Fender} R.~P.,  {Homan} J.,   {Belloni} T.~M.,  2009, \mn@doi [\mnras]
  {10.1111/j.1365-2966.2009.14841.x}, \href
  {https://ui.adsabs.harvard.edu/abs/2009MNRAS.396.1370F} {396, 1370}

\bibitem[\protect\citeauthoryear{{Fender} et~al.,}{{Fender}
  et~al.}{2016}]{Fender2017}
{Fender} R.,  et~al., 2016, in MeerKAT Science: On the Pathway to the SKA.
  p.~13 (\mn@eprint {arXiv} {1711.04132})

\bibitem[\protect\citeauthoryear{{Fender}, {Bright}, {Mooley}  \&
  {Miller-Jones}}{{Fender} et~al.}{2019}]{2019MNRAS.490L..76F}
{Fender} R.,  {Bright} J.,  {Mooley} K.,   {Miller-Jones} J.,  2019, \mn@doi
  [\mnras] {10.1093/mnrasl/slz145}, \href
  {https://ui.adsabs.harvard.edu/abs/2019MNRAS.490L..76F} {490, L76}

\bibitem[\protect\citeauthoryear{{Fomalont}, {Geldzahler}  \&
  {Bradshaw}}{{Fomalont} et~al.}{2001}]{2001ApJ...558..283F}
{Fomalont} E.~B.,  {Geldzahler} B.~J.,   {Bradshaw} C.~F.,  2001, \mn@doi
  [\apj] {10.1086/322479}, \href
  {https://ui.adsabs.harvard.edu/abs/2001ApJ...558..283F} {558, 283}

\bibitem[\protect\citeauthoryear{{Gallo}, {Degenaar}  \& {van den
  Eijnden}}{{Gallo} et~al.}{2018}]{2018MNRAS.478L.132G}
{Gallo} E.,  {Degenaar} N.,   {van den Eijnden} J.,  2018, \mn@doi [\mnras]
  {10.1093/mnrasl/sly083}, \href
  {https://ui.adsabs.harvard.edu/abs/2018MNRAS.478L.132G} {478, L132}

\bibitem[\protect\citeauthoryear{{Gusinskaia} et~al.,}{{Gusinskaia}
  et~al.}{2017}]{2017MNRAS.470.1871G}
{Gusinskaia} N.~V.,  et~al., 2017, \mn@doi [\mnras] {10.1093/mnras/stx1235},
  \href {https://ui.adsabs.harvard.edu/abs/2017MNRAS.470.1871G} {470, 1871}

\bibitem[\protect\citeauthoryear{{Hare} et~al.,}{{Hare}
  et~al.}{2020}]{2020ApJ...890...57H}
{Hare} J.,  et~al., 2020, \mn@doi [\apj] {10.3847/1538-4357/ab6a12}, \href
  {https://ui.adsabs.harvard.edu/abs/2020ApJ...890...57H} {890, 57}

\bibitem[\protect\citeauthoryear{{Hasinger} \& {van der Klis}}{{Hasinger} \&
  {van der Klis}}{1989}]{1989A&A...225...79H}
{Hasinger} G.,  {van der Klis} M.,  1989, \aap, \href
  {https://ui.adsabs.harvard.edu/abs/1989A&A...225...79H} {225, 79}

\bibitem[\protect\citeauthoryear{{Heywood}}{{Heywood}}{2020}]{2020ascl.soft09003H}
{Heywood} I.,  2020, {oxkat: Semi-automated imaging of MeerKAT observations}
  (\mn@eprint {ascl} {2009.003})

\bibitem[\protect\citeauthoryear{{Hickish} et~al.,}{{Hickish}
  et~al.}{2018}]{2018MNRAS.475.5677H}
{Hickish} J.,  et~al., 2018, \mn@doi [\mnras] {10.1093/mnras/sty074}, \href
  {https://ui.adsabs.harvard.edu/abs/2018MNRAS.475.5677H} {475, 5677}

\bibitem[\protect\citeauthoryear{{Hjellming} \& {Rupen}}{{Hjellming} \&
  {Rupen}}{1995}]{1995Natur.375..464H}
{Hjellming} R.~M.,  {Rupen} M.~P.,  1995, \mn@doi [\nat] {10.1038/375464a0},
  \href {https://ui.adsabs.harvard.edu/abs/1995Natur.375..464H} {375, 464}

\bibitem[\protect\citeauthoryear{{Homan} et~al.,}{{Homan}
  et~al.}{2006}]{2006ATel..725....1H}
{Homan} J.,  et~al., 2006, The Astronomer's Telegram, \href
  {https://ui.adsabs.harvard.edu/abs/2006ATel..725....1H} {725, 1}

\bibitem[\protect\citeauthoryear{{Homan} et~al.,}{{Homan}
  et~al.}{2007}]{2007ApJ...656..420H}
{Homan} J.,  et~al., 2007, \mn@doi [\apj] {10.1086/510447}, \href
  {https://ui.adsabs.harvard.edu/abs/2007ApJ...656..420H} {656, 420}

\bibitem[\protect\citeauthoryear{{Homan} et~al.,}{{Homan}
  et~al.}{2010}]{2010ApJ...719..201H}
{Homan} J.,  et~al., 2010, \mn@doi [\apj] {10.1088/0004-637X/719/1/201}, \href
  {https://ui.adsabs.harvard.edu/abs/2010ApJ...719..201H} {719, 201}

\bibitem[\protect\citeauthoryear{{Kennea} \& {Krimm}}{{Kennea} \&
  {Krimm}}{2018a}]{Kennea2018}
{Kennea} J.~A.,  {Krimm} H.~A.,  2018a, The Astronomer's Telegram, \href
  {https://ui.adsabs.harvard.edu/abs/2018ATel12160....1K} {12160}

\bibitem[\protect\citeauthoryear{{Kennea} \& {Krimm}}{{Kennea} \&
  {Krimm}}{2018b}]{2018ATel12160....1K}
{Kennea} J.~A.,  {Krimm} H.~A.,  2018b, The Astronomer's Telegram, \href
  {https://ui.adsabs.harvard.edu/abs/2018ATel12160....1K} {12160, 1}

\bibitem[\protect\citeauthoryear{Krimm et~al.,}{Krimm et~al.}{2013}]{Krimm2013}
Krimm H.~A.,  et~al., 2013, \mn@doi [The Astrophysical Journal Supplement
  Series] {10.1088/0067-0049/209/1/14}, 209, 14

\bibitem[\protect\citeauthoryear{{Krimm} et~al.,}{{Krimm}
  et~al.}{2018}]{Krimm2018}
{Krimm} H.~A.,  et~al., 2018, The Astronomer's Telegram, \href
  {https://ui.adsabs.harvard.edu/abs/2018ATel12151....1K} {12151}

\bibitem[\protect\citeauthoryear{{Lasota}}{{Lasota}}{2001}]{2001NewAR..45..449L}
{Lasota} J.-P.,  2001, \mn@doi [\nar] {10.1016/S1387-6473(01)00112-9}, \href
  {https://ui.adsabs.harvard.edu/abs/2001NewAR..45..449L} {45, 449}

\bibitem[\protect\citeauthoryear{{Lewin}, {Vacca}  \& {Basinska}}{{Lewin}
  et~al.}{1984}]{1984ApJ...277L..57L}
{Lewin} W.~H.~G.,  {Vacca} W.~D.,   {Basinska} E.~M.,  1984, \mn@doi [\apjl]
  {10.1086/184202}, \href
  {https://ui.adsabs.harvard.edu/abs/1984ApJ...277L..57L} {277, L57}

\bibitem[\protect\citeauthoryear{{Ludlam} et~al.,}{{Ludlam}
  et~al.}{2018}]{2018ATel12158....1L}
{Ludlam} R.~M.,  et~al., 2018, The Astronomer's Telegram, \href
  {https://ui.adsabs.harvard.edu/abs/2018ATel12158....1L} {12158, 1}

\bibitem[\protect\citeauthoryear{{Ludlam} et~al.,}{{Ludlam}
  et~al.}{2019}]{2019ApJ...873...99L}
{Ludlam} R.~M.,  et~al., 2019, \mn@doi [\apj] {10.3847/1538-4357/ab0414}, \href
  {https://ui.adsabs.harvard.edu/abs/2019ApJ...873...99L} {873, 99}

\bibitem[\protect\citeauthoryear{{Matsuoka} et~al.,}{{Matsuoka}
  et~al.}{2009}]{2009PASJ...61..999M}
{Matsuoka} M.,  et~al., 2009, \mn@doi [\pasj] {10.1093/pasj/61.5.999}, \href
  {https://ui.adsabs.harvard.edu/abs/2009PASJ...61..999M} {61, 999}

\bibitem[\protect\citeauthoryear{{Mazzola} et~al.,}{{Mazzola}
  et~al.}{2021}]{2021arXiv210800729M}
{Mazzola} S.~M.,  et~al., 2021, arXiv e-prints, \href
  {https://ui.adsabs.harvard.edu/abs/2021arXiv210800729M} {p. arXiv:2108.00729}

\bibitem[\protect\citeauthoryear{{McMullin}, {Waters}, {Schiebel}, {Young}  \&
  {Golap}}{{McMullin} et~al.}{2007}]{2007ASPC..376..127M}
{McMullin} J.~P.,  {Waters} B.,  {Schiebel} D.,  {Young} W.,   {Golap} K.,
  2007, in {Shaw} R.~A.,  {Hill} F.,   {Bell} D.~J.,  eds,  Astronomical
  Society of the Pacific Conference Series Vol. 376, Astronomical Data Analysis
  Software and Systems XVI. p.~127

\bibitem[\protect\citeauthoryear{{Migliari} \& {Fender}}{{Migliari} \&
  {Fender}}{2006}]{2006MNRAS.366...79M}
{Migliari} S.,  {Fender} R.~P.,  2006, \mn@doi [\mnras]
  {10.1111/j.1365-2966.2005.09777.x}, \href
  {https://ui.adsabs.harvard.edu/abs/2006MNRAS.366...79M} {366, 79}

\bibitem[\protect\citeauthoryear{{Migliari}, {Fender}, {Rupen}, {Jonker},
  {Klein-Wolt}, {Hjellming}  \& {van der Klis}}{{Migliari}
  et~al.}{2003}]{2003MNRAS.342L..67M}
{Migliari} S.,  {Fender} R.~P.,  {Rupen} M.,  {Jonker} P.~G.,  {Klein-Wolt} M.,
   {Hjellming} R.~M.,   {van der Klis} M.,  2003, \mn@doi [\mnras]
  {10.1046/j.1365-8711.2003.06795.x}, \href
  {https://ui.adsabs.harvard.edu/abs/2003MNRAS.342L..67M} {342, L67}

\bibitem[\protect\citeauthoryear{{Migliari}, {Fender}, {Rupen}, {Wachter},
  {Jonker}, {Homan}  \& {van der Klis}}{{Migliari}
  et~al.}{2004}]{2004MNRAS.351..186M}
{Migliari} S.,  {Fender} R.~P.,  {Rupen} M.,  {Wachter} S.,  {Jonker} P.~G.,
  {Homan} J.,   {van der Klis} M.,  2004, \mn@doi [\mnras]
  {10.1111/j.1365-2966.2004.07768.x}, \href
  {https://ui.adsabs.harvard.edu/abs/2004MNRAS.351..186M} {351, 186}

\bibitem[\protect\citeauthoryear{{Miller-Jones} et~al.,}{{Miller-Jones}
  et~al.}{2010}]{2010ApJ...716L.109M}
{Miller-Jones} J.~C.~A.,  et~al., 2010, \mn@doi [\apjl]
  {10.1088/2041-8205/716/2/L109}, \href
  {https://ui.adsabs.harvard.edu/abs/2010ApJ...716L.109M} {716, L109}

\bibitem[\protect\citeauthoryear{{Mirabel} \& {Rodr{\'\i}guez}}{{Mirabel} \&
  {Rodr{\'\i}guez}}{1994}]{1994Natur.371...46M}
{Mirabel} I.~F.,  {Rodr{\'\i}guez} L.~F.,  1994, \mn@doi [\nat]
  {10.1038/371046a0}, \href
  {https://ui.adsabs.harvard.edu/abs/1994Natur.371...46M} {371, 46}

\bibitem[\protect\citeauthoryear{{Monageng}, {Motta}, {Fender}, {Yu}, {Woudt},
  {Tremou}, {Miller-Jones}  \& {van der Horst}}{{Monageng}
  et~al.}{2021}]{2021MNRAS.501.5776M}
{Monageng} I.~M.,  {Motta} S.~E.,  {Fender} R.,  {Yu} W.,  {Woudt} P.~A.,
  {Tremou} E.,  {Miller-Jones} J.~C.~A.,   {van der Horst} A.~J.,  2021,
  \mn@doi [\mnras] {10.1093/mnras/stab043}, \href
  {https://ui.adsabs.harvard.edu/abs/2021MNRAS.501.5776M} {501, 5776}

\bibitem[\protect\citeauthoryear{{Mu{\~n}oz-Darias}, {Fender}, {Motta}  \&
  {Belloni}}{{Mu{\~n}oz-Darias} et~al.}{2014}]{2014MNRAS.443.3270M}
{Mu{\~n}oz-Darias} T.,  {Fender} R.~P.,  {Motta} S.~E.,   {Belloni} T.~M.,
  2014, \mn@doi [\mnras] {10.1093/mnras/stu1334}, \href
  {https://ui.adsabs.harvard.edu/abs/2014MNRAS.443.3270M} {443, 3270}

\bibitem[\protect\citeauthoryear{{Mu{\~n}oz-Darias}, {Jimenez-Ibarra}, {Armas
  Padilla}, {Casares}, {Cuneo}, {Panizo-Espinar}, {Sanchez-Sierras}  \&
  {Torres}}{{Mu{\~n}oz-Darias} et~al.}{2019}]{2019ATel12881....1M}
{Mu{\~n}oz-Darias} T.,  {Jimenez-Ibarra} F.,  {Armas Padilla} M.,  {Casares}
  J.,  {Cuneo} V.,  {Panizo-Espinar} G.,  {Sanchez-Sierras} J.,   {Torres}
  M.~A.~P.,  2019, The Astronomer's Telegram, \href
  {https://ui.adsabs.harvard.edu/abs/2019ATel12881....1M} {12881, 1}

\bibitem[\protect\citeauthoryear{{Mu{\~n}oz-Darias} et~al.,}{{Mu{\~n}oz-Darias}
  et~al.}{2020}]{2020ApJ...893L..19M}
{Mu{\~n}oz-Darias} T.,  et~al., 2020, \mn@doi [\apjl]
  {10.3847/2041-8213/ab8381}, \href
  {https://ui.adsabs.harvard.edu/abs/2020ApJ...893L..19M} {893, L19}

\bibitem[\protect\citeauthoryear{Narayan \& Yi}{Narayan \&
  Yi}{1995}]{Narayan1995}
Narayan R.,  Yi I.,  1995, The Astrophysical Journal, 452, 710

\bibitem[\protect\citeauthoryear{{Negoro} et~al.,}{{Negoro}
  et~al.}{2020}]{2020ATel13455....1N}
{Negoro} H.,  et~al., 2020, The Astronomer's Telegram, \href
  {https://ui.adsabs.harvard.edu/abs/2020ATel13455....1N} {13455, 1}

\bibitem[\protect\citeauthoryear{{Offringa} et~al.,}{{Offringa}
  et~al.}{2014}]{2014MNRAS.444..606O}
{Offringa} A.~R.,  et~al., 2014, \mn@doi [\mnras] {10.1093/mnras/stu1368},
  \href {https://ui.adsabs.harvard.edu/abs/2014MNRAS.444..606O} {444, 606}

\bibitem[\protect\citeauthoryear{{Paice}, {Gandhi}, {Dhillon}, {Marsh}, {Green}
   \& {Breedt}}{{Paice} et~al.}{2018}]{2018ATel12197....1P}
{Paice} J.~A.,  {Gandhi} P.,  {Dhillon} V.~S.,  {Marsh} T.~R.,  {Green} M.,
  {Breedt} E.,  2018, The Astronomer's Telegram, \href
  {https://ui.adsabs.harvard.edu/abs/2018ATel12197....1P} {12197, 1}

\bibitem[\protect\citeauthoryear{{Panurach} et~al.,}{{Panurach}
  et~al.}{2021}]{2021ApJ...923...88P}
{Panurach} T.,  et~al., 2021, \mn@doi [\apj] {10.3847/1538-4357/ac2c6b}, \href
  {https://ui.adsabs.harvard.edu/abs/2021ApJ...923...88P} {923, 88}

\bibitem[\protect\citeauthoryear{{Parikh}, {Wijnands}  \&
  {Altamirano}}{{Parikh} et~al.}{2020}]{2020ATel13725....1P}
{Parikh} A.~S.,  {Wijnands} R.,   {Altamirano} D.,  2020, The Astronomer's
  Telegram, \href {https://ui.adsabs.harvard.edu/abs/2020ATel13725....1P}
  {13725, 1}

\bibitem[\protect\citeauthoryear{{Perrott} et~al.,}{{Perrott}
  et~al.}{2013}]{2013MNRAS.429.3330P}
{Perrott} Y.~C.,  et~al., 2013, \mn@doi [\mnras] {10.1093/mnras/sts589}, \href
  {https://ui.adsabs.harvard.edu/abs/2013MNRAS.429.3330P} {429, 3330}

\bibitem[\protect\citeauthoryear{{Rajwade} et~al.,}{{Rajwade}
  et~al.}{2019}]{Rajwade2019}
{Rajwade} K.~M.,  et~al., 2019, The Astronomer's Telegram, \href
  {https://ui.adsabs.harvard.edu/abs/2019ATel12499....1R} {12499}

\bibitem[\protect\citeauthoryear{{Rana} et~al.,}{{Rana}
  et~al.}{2016}]{2016ApJ...821..103R}
{Rana} V.,  et~al., 2016, \mn@doi [\apj] {10.3847/0004-637X/821/2/103}, \href
  {https://ui.adsabs.harvard.edu/abs/2016ApJ...821..103R} {821, 103}

\bibitem[\protect\citeauthoryear{{Remillard} \& {McClintock}}{{Remillard} \&
  {McClintock}}{2006}]{2006ARA&A..44...49R}
{Remillard} R.~A.,  {McClintock} J.~E.,  2006, \mn@doi [\araa]
  {10.1146/annurev.astro.44.051905.092532}, \href
  {https://ui.adsabs.harvard.edu/abs/2006ARA&A..44...49R} {44, 49}

\bibitem[\protect\citeauthoryear{{Russell}, {Miller-Jones}, {Maccarone},
  {Yang}, {Fender}  \& {Lewis}}{{Russell} et~al.}{2011}]{2011ApJ...739L..19R}
{Russell} D.~M.,  {Miller-Jones} J.~C.~A.,  {Maccarone} T.~J.,  {Yang} Y.~J.,
  {Fender} R.~P.,   {Lewis} F.,  2011, \mn@doi [\apjl]
  {10.1088/2041-8205/739/1/L19}, \href
  {https://ui.adsabs.harvard.edu/abs/2011ApJ...739L..19R} {739, L19}

\bibitem[\protect\citeauthoryear{{Russell} et~al.,}{{Russell}
  et~al.}{2019}]{2019ApJ...883..198R}
{Russell} T.~D.,  et~al., 2019, \mn@doi [\apj] {10.3847/1538-4357/ab3d36},
  \href {https://ui.adsabs.harvard.edu/abs/2019ApJ...883..198R} {883, 198}

\bibitem[\protect\citeauthoryear{{Russell} et~al.,}{{Russell}
  et~al.}{2021}]{2021MNRAS.508L...6R}
{Russell} T.~D.,  et~al., 2021, \mn@doi [\mnras] {10.1093/mnrasl/slab087},
  \href {https://ui.adsabs.harvard.edu/abs/2021MNRAS.508L...6R} {508, L6}

\bibitem[\protect\citeauthoryear{{Saikia}, {Russell}, {Baglio}, {Bramich}  \&
  {Lewis}}{{Saikia} et~al.}{2020}]{2020ATel13719....1S}
{Saikia} P.,  {Russell} D.~M.,  {Baglio} M.~C.,  {Bramich} D.~M.,   {Lewis} F.,
   2020, The Astronomer's Telegram, \href
  {https://ui.adsabs.harvard.edu/abs/2020ATel13719....1S} {13719, 1}

\bibitem[\protect\citeauthoryear{{Tawara} et~al.,}{{Tawara}
  et~al.}{1984}]{1984ApJ...276L..41T}
{Tawara} Y.,  et~al., 1984, \mn@doi [\apjl] {10.1086/184184}, \href
  {https://ui.adsabs.harvard.edu/abs/1984ApJ...276L..41T} {276, L41}

\bibitem[\protect\citeauthoryear{{Zwart} et~al.,}{{Zwart}
  et~al.}{2008}]{2008MNRAS.391.1545Z}
{Zwart} J.~T.~L.,  et~al., 2008, \mn@doi [\mnras]
  {10.1111/j.1365-2966.2008.13953.x}, \href
  {https://ui.adsabs.harvard.edu/abs/2008MNRAS.391.1545Z} {391, 1545}

\bibitem[\protect\citeauthoryear{{van den Eijnden} et~al.,}{{van den Eijnden}
  et~al.}{2020}]{2020MNRAS.496.4127V}
{van den Eijnden} J.,  et~al., 2020, \mn@doi [\mnras] {10.1093/mnras/staa1704},
  \href {https://ui.adsabs.harvard.edu/abs/2020MNRAS.496.4127V} {496, 4127}

\bibitem[\protect\citeauthoryear{{van der Klis}}{{van der
  Klis}}{1989}]{1989ESASP.296..203V}
{van der Klis} M.,  1989, in {Hunt} J.,  {Battrick} B.,  eds,  ESA Special
  Publication Vol. 1, Two Topics in X-Ray Astronomy, Volume 1: X Ray Binaries.
  Volume 2: AGN and the X Ray Background. p.~203

\bibitem[\protect\citeauthoryear{{van der Klis}}{{van der
  Klis}}{1994}]{1994ApJS...92..511V}
{van der Klis} M.,  1994, \mn@doi [\apjs] {10.1086/192006}, \href
  {https://ui.adsabs.harvard.edu/abs/1994ApJS...92..511V} {92, 511}

\makeatother
\end{thebibliography}



\appendix

\section{Radio fluxes}

\begin{table}
\caption{List of observations made with AMI-LA with 15.5\,GHz. Each with date, start time, the flux density and duration. The uncertainties quoted are the statistical error and a 5\% calibration uncertainty added in quadrature. For epochs where we did not detect J1858, we provide a 3$\sigma$ upper limit with the prefix `$<$'.}
\label{tab:AMI_obs}
\begin{tabular}{ccc}
\hline
Start time (MJD) & Flux density ($\mu$Jy/beam) & Duration (hrs) \\
\hline
58424.574 & 310$\pm$50 & 2 \\
58425.712 & 390$\pm$40 & 2 \\
58427.546 & 590$\pm$60 & 4 \\
58430.605 & 390$\pm$40 & 2 \\
58440.540 & 360$\pm$30 & 3 \\
58454.537 & 300$\pm$40 & 3 \\
58464.501 & 330$\pm$20 & 4 \\
58468.440 & 360$\pm$30 & 4 \\
58475.421 & $<$200 & 4 \\
58482.523 & 310$\pm$30 & 4 \\
58489.385 & 180$\pm$40 & 4 \\
58496.363 & 280$\pm$20 & 4 \\
58503.344 & 300$\pm$30 & 4 \\
58513.350 & $<$170 & 4 \\
58517.185 & 340$\pm$40 & 4 \\
58524.270 & $<$230 & 4 \\
58531.266 & 370$\pm$30 & 6 \\
58538.199 & 160$\pm$10 & 6 \\
58544.306 & $<$130 & 6 \\
58559.144 & 200$\pm$20 & 6 \\
58565.211 & $<$110 & 6 \\
58587.060 & 180$\pm$20 & 5 \\
58606.013 & $<$270 & 6 \\
58614.026 & $<$230 & 6 \\
58620.983 & $<$450 & 6 \\
58634.955 & $<$230 & 8 \\
58640.913 & $<$140 & 6 \\
58648.928 & $<$140 & 6 \\
58655.872 & $<$370 & 7 \\
58662.853 & $<$170 & 7 \\
58669.871 & $<$220 & 6 \\
58676.818 & $<$250 & 8 \\
58686.949 & 160$\pm$10 & 6 \\
58697.844 & 260$\pm$10 & 4 \\
58700.836 & 200$\pm$20 & 4 \\
58701.852 & 300$\pm$20 & 3.5 \\
58711.846 & $<$340 & 3 \\
58725.773 & 220$\pm$20 & 4 \\
58729.189 & 180$\pm$10 & 6 \\
58732.748 & 240$\pm$20 & 3.8 \\
58739.741 & 280$\pm$20 & 4 \\
58746.710 & 200$\pm$10 & 4 \\
58760.672 & 160$\pm$10 & 4 \\
58775.631 & $<$140 & 4 \\
58781.614 & $<$360 & 1 \\
58795.576 & 240$\pm$30 & 4 \\
58802.557 & 170$\pm$10 & 4 \\
58809.538 & $<$190 & 4 \\
58816.519 & 120$\pm$10 & 4\\
58822.500 & 350$\pm$60 & 2 \\
58851.423 & $<$230 & 4  \\
58858.404 & $<$140 & 4 \\
58867.380 & 160$\pm$20 & 4\\
58872.366 & 290$\pm$20 & 4 \\
58879.350 & 370$\pm$50 & 4 \\
58886.330 & 200$\pm$10 & 4 \\
58911.260 & $<$290 & 4\\
58914.251 & $<$180 & 4 \\
58924.224 & $<$200 & 4\\
\hline
\end{tabular}
\end{table}

\begin{table}
 \caption{List of observations made with MeerKAT at 1.4\,GHz. Each with date, start time, the flux density and duration. The uncertainties are calculated by adding the statistical and 10\% calibration uncertainty in quadrature. On occasions where the source was not detected we provide a 3$\sigma$ upper limit with the prefix `$<$'.}
 \label{tab:MeerKAT_obs}
 \begin{tabular}{ccc}
  \hline
   Start Time (MJD) & Flux density ($\mu$Jy/beam) & Duration (hrs) \\
  \hline
58432.505 & 126$\pm$20& 0.25 \\
58439.491 & 84$\pm$12 & 0.25 \\
58446.458 & 114$\pm$14 & 0.25 \\
58454.434 & 132$\pm$15 & 0.25 \\
58460.416 & 126$\pm$17 & 0.25 \\
58467.626 & 101$\pm$13 & 0.25 \\
58474.592 & 62$\pm$14 & 0.25 \\
58481.571 & $<$57 & 0.25 \\
58488.107 & 119$\pm$18 & 0.25 \\
58530.178 & 993$\pm$101 & 0.25 \\
58543.126 & 102$\pm$17 & 0.25 \\
58551.081 & 76$\pm$9 & 0.25 \\
58560.121 & $<$60 & 0.25 \\
58567.121 & $<$48 & 0.25 \\
58574.107 & 85$\pm$15 & 0.25 \\
58582.101 & 94$\pm$10 & 0.25 \\
58588.101 & $<$57 & 0.25 \\
58593.123 & 57$\pm$9 & 0.25 \\
58602.193 & $<$57 & 0.25 \\
58607.997 & 62$\pm$7 & 0.25 \\
58614.942 & $<$63 & 0.25 \\
58621.931 & $<$57 & 0.25 \\
58626.053 & $<$54 & 0.25 \\
58700.766 & 83$\pm$9 & 1.0 \\
58910.193 & $<$63 & 0.25 \\
  \hline
 \end{tabular}
\end{table}

\begin{table}
    \centering
        \caption{List of the spectral index calculations of both the new data presented in this table as well as epochs used from \citet{2020MNRAS.496.4127V}. This data is shown in the second panel of Figure \ref{fig:lc}. }
    \begin{tabular}{ccc}
\hline
Date (MJD)	&	Spectral Index			&	Frequency Range (GHz)	\\
\hline
58530.178	&	0.9	$\pm$	0.5	&	1.0-1.6	\\
58432.227	&	1.1	$\pm$	0.1	&	1.3-4.5	\\
58439.168	&	1.1	$\pm$	0.1	&	1.3-4.5 \\
58700.905	&	0.6	$\pm$	0.1	&	1.3-4.5 \\
58566.800	&	0.8	$\pm$	0.1	&	1.3-4.5 \\
58431.954	&	0.4	$\pm$	0.2	&	4.5-7.5	\\
58436.072	&	0.3	$\pm$	0.2	&	4.5-7.5	\\
58437.060	&	0.1	$\pm$	0.1	&	4.5-7.5	\\
58438.835	&	0.6	$\pm$	0.2	&	4.5-7.5	\\
58443.748	&	0.6	$\pm$	0.3	&	4.5-7.5	\\
58443.922	&	0.6	$\pm$	0.3	&	4.5-7.5	\\
58534.498	&	0.0	$\pm$	0.2	&	4.5-7.5	\\
58566.479	&	0.7	$\pm$	0.3	&	4.5-7.5	\\
58701.035	&	0.1	$\pm$	0.1	&	4.5-7.5	\\
58430.207	&	0.4	$\pm$	0.1	&	5.5-9	\\
58430.405	&	-0.1$\pm$	0.1	&	5.5-15.5\\
58439.491	&	0.6	$\pm$	0.1	&	1.3-15.5\\
58454.434	&	0.3	$\pm$	0.1	&	1.3-15.5\\
58467.630	&	0.5	$\pm$	0.1	&	1.3-15.5\\
58474.592	&	0.7	$\pm$	0.1	&	1.3-15.5\\
58488.508	&	0.2	$\pm$	0.1	&	1.3-15.5\\
58543.126	&	0.0	$\pm$	0.1	&	1.3-15.5\\
58607.997	&	0.8	$\pm$	0.1	&	1.3-15.5\\
58560.121	&	0.5	$\pm$	0.1	&	1.3-15.5\\
58614.956	&	0.7	$\pm$	0.1	&	1.3-15.5\\
58700.776	&	0.5	$\pm$	0.1	&	1.3-15.5\\

\hline
    \end{tabular}
    \label{tab:spectral_index}
\end{table}



\bsp	
\label{lastpage}
\end{document}